\input harvmac

\def \lg {{\rm log}}

\def \e {{\rm e}}
\def \ov {\over}

\def \F {{\cal F}}\def \ep {\epsilon}
\def \k {\kappa}
\def \N {{\cal N}}
\def \L {{\cal L}}

\def \pa { \partial}
\def \a {\alpha}

\def \g {\gamma}
\def \G {\Gamma}
\def \d {\delta}
\def \D {\Delta}
\def \l {\lambda}
\def \La {\Lambda}

\def \ee {\epsilon}

\def \p {\phi}

\def \tx {\tilde x}
\def \tS{\tilde S}

\def \bt {\bar{t}}
\def \bN {\overline{N}}
\def \bC {\overline{C}}

\def \dXL {\dot{X}_L}
\def \dXR {\dot{X}_R}
\def \rL   {{\rm L}}

\def \lr { \lref}

\def \lr{\lref}

\def \rf {\refs}

\lref\KW{I.R. Klebanov and E. Witten,
``Superconformal Field Theory on
Threebranes at a Calabi-Yau Singularity,''
Nucl.\ Phys.\ B {\bf 536}, 199 (1998)
[hep-th/9807080].
}

\lr\rom{L.~J.~Romans,
``New Compactifications Of Chiral N=2 D = 10 Supergravity,''
Phys.\ Lett.\ B {\bf 153}, 392 (1985).
}

\lref\mar{D.~Marolf,
``Chern-Simons terms and the three notions of charge,''
hep-th/0006117.
}

\lref\gug{
M.~B.~Green and M.~Gutperle,
``Effects of D-instantons,''
Nucl.\ Phys.\ B {\bf 498}, 195 (1997)
[hep-th/9701093].
I.~Antoniadis, S.~Ferrara, R.~Minasian and K.~S.~Narain,
``$R^4$ couplings in M- and type II theories on Calabi-Yau spaces,''
Nucl.\ Phys.\ B {\bf 507}, 571 (1997)
[hep-th/9707013].
E.~Kiritsis and B.~Pioline,
``On $R^4$ threshold corrections in type IIB string theory and (p,q) string  instantons,''
Nucl.\ Phys.\ B {\bf 508}, 509 (1997)
[hep-th/9707018].
}

\lref\KS{I.~R.~Klebanov and M.~J.~Strassler,
``Supergravity and a Confining Gauge Theory:
Duality Cascades and $\chi$SB-Resolution of Naked Singularities,''
JHEP {\bf 0008}, 052 (2000)
[hep-th/0007191].
}

\lr\gip{
S.~B.~Giddings, S.~Kachru and J.~Polchinski,
``Hierarchies from fluxes in string compactifications,''
hep-th/0105097.
}

\lr\TTT{
A.~A.~Tseytlin,
``$R^4$ terms in 11 dimensions and
conformal anomaly of (2,0) theory,''
Nucl.\ Phys.\ B {\bf 584}, 233 (2000)
[hep-th/0005072].
}

\lr\BB{
K.~Becker and M.~Becker,
``Supersymmetry breaking, M-theory and fluxes,''
JHEP {\bf 0107}, 038 (2001)
[hep-th/0107044].
}

\lref\GK{
S. S. Gubser and I. R. Klebanov, ``Baryons and Domain Walls in an
{\cal N}=1 Superconformal Gauge Theory,'' {Phys. Rev.} {\bf D58}
 125025 (1998), [hep-th/9808075].
}

\lref\NT{
B.E.W. Nilsson and A.K. Tollsten,
``Supersymmetrization of $\zeta(3) R^4$ in superstring theories,"
Phys.\ Lett.\ B {\bf 181}, 63 (1986).
  R. Kallosh, ``Strings And Superspace,''
Phys.\ Scripta {\bf T15}, 118 (1987).
}

\lref\MP{
D.~R.~Morrison and M.~R.~Plesser,
``Non-spherical horizons. I,''
Adv.\ Theor.\ Math.\ Phys.\  {\bf 3}, 1 (1999)
[hep-th/9810201].
}

\lref\HKO{C.P. Herzog, I.R. Klebanov and P. Ouyang,
``Remarks on the Warped Deformed Conifold,''
{\tt hep-th/0108101}.
}

\lref\KN{
I.~R.~Klebanov and N.~Nekrasov,
``Gravity Duals of Fractional Branes and Logarithmic RG Flow,''
{Nucl. Phys.} {\bf B574} (2000) 263,
{\tt hep-th/9911096}.
}

\lref\Witten{
E.~Witten,
``Phases of N = 2 theories in two dimensions,''
Nucl.\ Phys.\ B {\bf 403}, 159 (1993)
[hep-th/9301042].
}

\lr\mtt{
R.~R.~Metsaev and A.~A.~Tseytlin,
``Order alpha prime (two loop) equivalence ef
 the string equations   of motion and the  sigma model
 Weyl invariance conditions: dependence on  the
dilaton and the antisymmetric tensor,''
Nucl.\ Phys.\ B {\bf 293}, 385 (1987).
}

\lref\KR{R.~Kallosh and A.~Rajaraman,
``Vacua of M-theory and string theory,''
Phys.\ Rev.\ D {\bf 58}, 125003 (1998)
[hep-th/9805041].
}

\lref\gs{
D.~J.~Gross and J.~H.~Sloan,
``The Quartic Effective Action For The Heterotic String,''
Nucl.\ Phys.\ B {\bf 291}, 41 (1987).
}
\lref\pvw{
K.~Peeters, P.~Vanhove and A.~Westerberg,
``Supersymmetric higher-derivative actions in
ten and eleven dimensions,  the associated
superalgebras and their formulation in superspace,''
Class.\ Quant.\ Grav.\  {\bf 18}, 843 (2001)
[hep-th/0010167].
}
\lref\schwarz{
J.~H.~Schwarz,
``Superstring Theory,''
Phys.\ Rept.\  {\bf 89}, 223 (1982).
}
\lref\bhkzt{
A.~Buchel, C.~P.~Herzog, I.~R.~Klebanov, L.~Pando Zayas
and
A.~A.~Tseytlin,
``Non-extremal gravity duals for fractional D3-branes on
the conifold,'' JHEP {\bf 0104}, 033 (2001)
[hep-th/0102105].
S.~S.~Gubser, C.~P.~Herzog, I.~R.~Klebanov and A.~A.~Tseytlin,
``Restoration of Chiral Symmetry: A Supergravity Perspective,''
JHEP {\bf 0105}, 028 (2001)
[hep-th/0102172].
}
\lref\ts{
A.~A.~Tseytlin,
``Ambiguity In The Effective Action In String Theories,''
Phys.\ Lett.\ B {\bf 176}, 92 (1986).
}
\lref\kp{
A.~Kehagias and H.~Partouche,
``The exact quartic effective action for the type IIB superstring,''
Phys.\ Lett.\ B {\bf 422}, 109 (1998)
[hep-th/9710023].
}

\lref\gz{
M.~T.~Grisaru and D.~Zanon,
``Sigma Model Superstring Corrections To
The Einstein-Hilbert Action,''
Phys.\ Lett.\ B {\bf 177}, 347 (1986).
M.~D.~Freeman, C.~N.~Pope, M.~F.~Sohnius and K.~S.~Stelle,
``Higher Order Sigma Model Counterterms And
The Effective Action For Superstrings,''
Phys.\ Lett.\ B {\bf 178}, 199 (1986).
Q.-H.~Park and D.~Zanon,
``More On Sigma Model Beta Functions
And Low-Energy Effective Actions,''
Phys.\ Rev.\ D {\bf 35}, 4038 (1987).
}

\lref\grw{
D.~J.~Gross and E.~Witten,
``Superstring Modifications Of Einstein's Equations,''
Nucl.\ Phys.\ B {\bf 277}, 1 (1986).
}

\lr\bg{
T.~Banks and M.~B.~Green,
``Non-perturbative effects in
$AdS_5 \times S^5 $
string theory and d = 4 SUSY  Yang-Mills,''
JHEP {\bf 9805}, 002 (1998)
[hep-th/9804170].
}

\lr\mye{
R.~Myers,
``Superstring Gravity And Black Holes,''
Nucl.\ Phys.\ B {\bf 289}, 701 (1987).
}

\lr\dk{
A.~A.~Deriglazov and S.~V.~Ketov,
``Four loop divergences of the two-dimensional
(1,1) supersymmetric nonlinear sigma model
with a Wess-Zumino-Witten term,''
Nucl.\ Phys.\ B {\bf 359}, 498 (1991).
}

\lr\gkt{
S.~S.~Gubser, I.~R.~Klebanov and A.~A.~Tseytlin,
``Coupling constant dependence in the
thermodynamics of N = 4  supersymmetric Yang-Mills theory,''
Nucl.\ Phys.\ B {\bf 534}, 202 (1998)
[hep-th/9805156].
}

\lr\roo{
M.~de Roo, H.~Suelmann and A.~Wiedemann,
``The Supersymmetric effective action of
the heterotic string in ten-dimensions,''
Nucl.\ Phys.\ B {\bf 405}, 326 (1993)
[hep-th/9210099].
J.~H.~Suelmann,
``Supersymmetry and string effective actions,''
Ph.D. Thesis, Groningen, 1994, RX-1510.
}

\lr\gwz{
M.~T.~Grisaru, A.~E.~van de Ven and D.~Zanon,
``Four Loop Beta Function For The N=1 And N=2
Supersymmetric Nonlinear Sigma Model In Two-Dimensions,''
Phys.\ Lett.\ B {\bf 173}, 423 (1986).
``Two-Dimensional Supersymmetric Sigma Models
On Ricci Flat Kahler Manifolds Are Not Finite,''
Nucl.\ Phys.\ B {\bf 277}, 388 (1986).
}

\lref\KW{I.~R.~Klebanov and E.~Witten,
``Superconformal field theory on threebranes
at a Calabi-Yau  singularity,''
Nucl.\ Phys.\ B {\bf 536}, 199 (1998)
[hep-th/9807080].
}

\lref\MP{
D.~R.~Morrison and M.~R.~Plesser,
``Non-spherical horizons. I,''
Adv.\ Theor.\ Math.\ Phys.\ {\bf 3}, 1 (1999)
[hep-th/9810201].
}

\lref\GKP{
S.~S.~Gubser, I.~R.~Klebanov and A.~M.~Polyakov,
``Gauge theory correlators from non-critical string theory,''
Phys.\ Lett.\ B {\bf 428}, 105 (1998)
[hep-th/9802109].
}

\lref\EW{
E.~Witten,
``Anti-de Sitter space and holography,''
Adv.\ Theor.\ Math.\ Phys.\ {\bf 2}, 253 (1998)
[hep-th/9802150].
}

\lref \KN{
I.~R.~Klebanov and N.~A.~Nekrasov,
``Gravity duals of fractional branes and logarithmic RG flow,''
Nucl.\ Phys.\ B {\bf 574}, 263 (2000)
[hep-th/9911096].
}

\lr\pof{
M.~D.~Freeman and C.~N.~Pope,
``Beta Functions And Superstring Compactifications,''
Phys.\ Lett.\ B {\bf 174}, 48 (1986).
A.~Sen,
``Central Charge Of The Virasoro Algebra
For Supersymmetric Sigma Models On Calabi-Yau Manifolds,''
Phys.\ Lett.\ B {\bf 178}, 370 (1986).
}

\lref\KT{
I.~R.~Klebanov and A.~A.~Tseytlin,
``Gravity Duals of Supersymmetric
$SU(N) \times SU(N+M)$ Gauge Theories,''
{ Nucl. Phys.} {\bf B578}, 123 (2000),
[hep-th/0002159].
}

\lref\NS{
D.~Nemeschansky and A.~Sen,
``Conformal Invariance Of Supersymmetric Sigma Models
On Calabi-Yau Manifolds,''
Phys.\ Lett.\ B {\bf 178}, 365 (1986).
}
\lref\jjj{
I.~Jack,
``The Twisted N=2 supersymmetric sigma model: A Four loop calculation of the beta function,''
Nucl.\ Phys.\ B {\bf 371}, 482 (1992).
}

\lref\jthroat{
J.~Maldacena,
``The large N limit of superconformal
field theories and supergravity,''
Adv.\ Theor.\ Math.\ Phys.\ {\bf 2}, 231 (1998)
[hep-th/9711200].
 }

\lref\grsch{
M.~B.~Green and J.~H.~Schwarz,
``Supersymmetrical Dual String Theory,''
Nucl.\ Phys.\ B {\bf 181}, 502 (1981);
``Supersymmetrical Dual String Theory. 2. Vertices And Trees,''
Nucl.\ Phys.\ B {\bf 198}, 252 (1982).
}
\lref\gsw{
M.~B.~Green, J.~H.~Schwarz and E.~Witten,
``Superstring Theory. Vols. 1 and 2''
{\it  Cambridge Univ. Press ( 1987).}
}

\lref \kaz { M.~T.~Grisaru, D.~I.~Kazakov and D.~Zanon,
``Five Loop Divergences For The N=2 Supersymmetric
Nonlinear Sigma Model,''
Nucl.\ Phys.\ B {\bf 287}, 189 (1987).
}
\lref \mgr{
M.~B.~Green and S.~Sethi,
``Supersymmetry constraints on type IIB supergravity,''
Phys.\ Rev.\ D {\bf 59}, 046006 (1999)
[hep-th/9808061].
M.~B.~Green,
``Interconnections Between Type II Superstrings,
M Theory And N = 4  Yang-Mills,''
hep-th/9903124.
}
\lref \pwv{ K. Peeters, P. Vanhove and A. Westerberg,
to appear.}

\lref \shif{
V.~A.~Novikov, M.~A.~Shifman, A.~I.~Vainshtein and V.~I.~Zakharov,
``Supersymmetric Instanton Calculus: Gauge Theories With Matter,''
Nucl.\ Phys.\ B {\bf 260}, 157 (1985).
``Beta Function In Supersymmetric Gauge Theories: Instantons Versus Traditional Approach,''
Phys.\ Lett.\ B {\bf 166}, 329 (1986).
}
\lref \shiv{
M.~A.~Shifman and A.~I.~Vainshtein,
``Solution Of The Anomaly Puzzle In Susy Gauge Theories And The Wilson Operator Expansion,''
Nucl.\ Phys.\ B {\bf 277}, 456 (1986).
}

\lref\grs{
M.~B.~Green and N.~Seiberg,
``Contact Interactions In Superstring Theory,''
Nucl.\ Phys.\ B {\bf 299}, 559 (1988).
}

\Title{\vbox
{\baselineskip 10pt
{\hbox{OHSTPY-HEP-T-01-024}\hbox{PUPT-2004}
\hbox{hep-th/0108106}
}}}
{\vbox{\vskip -30 true pt\centerline {String Corrections to
the Holographic RG Flow of}
\medskip
\centerline {Supersymmetric  SU(N+M) x SU(N) Gauge Theory}
\medskip
\vskip4pt }}
\vskip -20 true pt
\centerline{Sergey Frolov$^1$\footnote{$^*$} {Also at Steklov
Mathematical Institute, Moscow.},
Igor R.~Klebanov$^{2}$ and
Arkady A.~Tseytlin$^{1}$\footnote{$^{**}$}
{Also at Imperial College, London and
 Lebedev Physics Institute, Moscow.}
 }
\smallskip\smallskip
\centerline{$^{1}$ \it  Department of Physics,
The Ohio State University,
Columbus, OH 43210, USA}
\centerline{$^{2}$ \it Joseph Henry
Laboratories, Princeton University, Princeton, NJ 08544, USA}

\bigskip\bigskip
\centerline {\bf Abstract}
\baselineskip12pt
\noindent
\medskip
We study leading string corrections to the type IIB
supergravity solution dual to the
${\cal N}=1$ supersymmetric $SU(N+M)\times
SU(N)$ gauge theory coupled to bifundamental
chiral superfields $A_i, B_j$, $i,j=1,2$.
This solution was found in hep-th/0007191, and its asymptotic form
describing logarithmic RG flow was constructed in
hep-th/0002159. The leading  tree-level string correction
to the type IIB  string effective action is represented by
the invariant of the form $\a'{}^3 ( R^4 + ...)$.
Since the background contains 3-form
field strengths, we need to know  parts of this invariant
that depend on them.  By analyzing the 5-point superstring
scattering amplitudes
we show that only a few specific  $R^3 (H_3)^2$ and $R^3 (F_3)^2$
terms
are present in the effective action. Their contribution to
the holographic RG flow turns out to be of the same order as of
the $R^4$ terms. Using this fact we show that it is possible to have
agreement between the $\alpha'$-corrected radial dependence of the
supergravity fields  and the RG flow dictated by the
NSVZ beta functions in field theory.
The agreement with field theory
requires that the anomalous dimension
of the  operators $\Tr (A_i B_j)$ is corrected by a term of order
$(M/N)^4 \lambda^{-1/2}$ from its value $-{1 \over 2} $ found for $M=0$
($\lambda$ is the appropriate 't Hooft coupling which is assumed to be
strong).
\bigskip

\Date{8/01}

\noblackbox
\baselineskip 16pt plus 2pt minus 2pt

\newsec{Introduction}

Investigations of D-branes on conifolds have produced interesting
examples of gauge/string duality with ${\cal N}=1$ supersymmetry.
The first case to be considered involves a large number $N$
of D3-branes placed at the singularity of the conifold,
which is a Calabi-Yau cone described by the
equation
$\sum_{i=1}^4 z_i^2 = 0$
in ${\bf C}^4$.
The near-horizon geometry produced by the D3-branes is
$AdS_5\times T^{1,1}$
where $T^{1,1}=(SU(2)\times  SU(2))/U(1)$ is the base of the cone.
Type IIB string theory on this background is conjectured to be dual
to the IR limit of the gauge theory on
the stack of D3-branes, which is
the ${\cal N}=1$ supersymmetric $SU(N)\times SU(N)$ gauge theory
coupled to bifundamental chiral superfields $A_1, A_2, B_1, B_2$
\refs{\KW,\MP}.

This duality may be generalized by adding $M$ fractional D3-branes
(wrapped D5-branes) to the $N$ regular D3-branes at the apex of
the conifold \rf{\GK,\KN}.
The $SU(N+M)\times SU(N)$ gauge theory on such a stack is
dual to a more complicated solution of
type IIB supergravity \refs{\KT,\KS}.
As was realized in \KS, in order to consistently
extend the original singular solution of \KT\
to the small radius (IR) region,
it is necessary to deform the conifold:
$\sum_{i=1}^4 z_i^2 = \epsilon^2$.

The presence of fractional branes
destroys conformal invariance,
and this ${\cal N}=1$ gauge theory exhibits an intricate pattern of
RG flows. The inverse-squared gauge couplings $1/g_1^2$ and
$1/g_2^2$ flow logarithmically in
opposite directions until the coupling
of the bigger gauge group diverges. To continue past this point it
is necessary to apply Seiberg duality to the bigger gauge group \KS.
This transformation maps the original gauge
theory to essentially the same
theory with $N$ replaced by $N-M$. After this,  the pattern of
the flow repeats itself. Thus, the RG flow
involves a series of duality
transformations. At the bottom of this duality cascade one finds
a gauge theory which exhibits chiral
symmetry breaking and confinement \KS.

All these features of the RG flow are nicely encoded in the dual
supergravity background. In the UV (for large radius)
one finds a logarithmic
flow of $1/g_1^2 - 1/g_2^2$ \refs{\KN,\KT}. In fact,
it was recently shown that the coefficient of the logarithm found in
supergravity agrees exactly with the prediction of field theoretic
NSVZ \shif,\shiv\  beta functions \HKO.
 The reduction in the rank of the gauge
group due to repeated cascade steps
is reflected in the radial dependence of the 5-form flux \KT.
In the IR the cascade is terminated by
the deformation of the conifold
which is responsible for the
chiral symmetry breaking and confinement \KS.

In this paper we examine modifications of the supergravity solutions
found in \rf{\KW,\KT,\KS} by the leading stringy effects encoded in
the $O(\a'{}^3)$ corrections to the effective action.
The paper has the following structure.
In section 2 we examine the structure of the  $O(\a'{}^3)$ terms in
the type IIB string effective action. We pay special attention to
terms that depend on the 3-form field strength. In particular,
we show that certain $R^3(H_3)^2$ terms
($R$ is the curvature and $H_3$
 is the NS-NS 3-form), which
were expected to be present in earlier literature, do not appear.
The $R^3(H_3)^2$ terms that do appear in the action,
(as well as 
other terms in 
NS-NS sector)
turn out to contribute
at the same order as the $R^4$ terms when evaluated on
the KT \KT\ solution.
\foot{We are indebted to K. Peeters,
P. Vanhove and A. Westerberg
for drawing our attention to  a wrong statement concerning
these terms in
the first version of this paper and informing us  about their
covariant NSR computation of $R^3 (H_3)^2 $  terms in the
 type IIB string 1-loop
effective action  \pwv.
 }
The necessary 5-point amplitude (Green-Schwarz light-cone gauge)
calculation
is delegated to Appendix A.
We will not analyze other similar terms  involving RR fields
which are not explicitly known at present 
but will conjecture  that since they are part of the 
same superinvariant, 
  their contribution should be again 
the  same as  of $R^4$ terms.

In section 3 we show that the $\a'{}^3$
correction does not modify the
$AdS_5 \times T^{1,1}$ background.
This is in line with expectations from the AdS/CFT
correspondence \refs{\jthroat,\GKP,\EW}:
the dual $SU(N)\times SU(N)$ gauge theory is conformal
for all values of $N$ and for all values of the gauge couplings.
Thus, the radius of $AdS_5 \times T^{1,1}$, which is related to
$g_s N$,
must be a free parameter (a modulus of CFT)
not
only in the  supergravity approximation
 but also in the full
string theoretic treatment.

In section 4 we
recall the structure of the supergravity solution \KT\ describing
logarithmic RG flow in the dual
$SU(N+M)\times SU(N)$ gauge theory, and review its comparison
with the NSVZ $\beta$-functions.
In section 5 we
study the $\a'{}^3$ corrections to this solution (their detailed
analysis is presented in Appendix B).
 In contrast to the $AdS_5 \times T^{1,1}$ case, here
 the corrections modify the form of the solution.
In particular, the  dilaton,
which was constant in the supergravity solution,
acquires radial dependence due to the stringy effects.
This translates into the RG flow of
the sum of inverse-square couplings
$1/g_1^2 + 1/g_2^2$. {}From the field theory
point of view,  this running is due
to a correction to the anomalous dimension of the operator
$\Tr (A_i B_j)$.
For $M=0$ this anomalous dimension is equal to $-1/2$
\KW\ but turning on
$M$ is expected to correct it by an even power of $M/N$ \KS.
We will see that, if $M\ll N$ and if the `t Hooft couplings are large,
string theory  predicts that this correction is
of order $\left ({M\over N}\right )^4
\left ( {1\over N g_1^2} + {1\over N g_2^2}\right )^{1/2}
.$

\newsec{Structure of $O(\a'{}^3)$
 terms in type IIB superstring effective action}

In this section we recall the structure of $\a'{}^3$ corrections
to type IIB effective action.
The classical type IIB  supergravity
 action in the normalizations
we use has the following form
\eqn\twob{
S=  {1\over 2\kappa^2 }\int\ d^{10}x \sqrt{-g}\bigg[ R-{1\over
2}(\partial \phi)^2  - {1\over
2} e^{2 \p}g_s^2 (\partial C_0)^2 -
   {1 \ov 12} (e^{-\p} H^2_3 +  e^{\p}  g_s^2 F^2_3)
-  { 1 \ov  4 \cdot 5!} g^2_s F^2_5  + ... \bigg]  .}
Here  $\kappa = 8 \pi^{7/2} g_s \a'^2$ is the gravitational
constant,
$\phi$ is the dilaton,
$C_0$ is the R-R scalar,
$H_3\equiv H=dB_2$ is the NS-NS 3-form,
$F_3\equiv F=dC_2$ is the R-R 3-form, and $F_5$ is the R-R
self-dual 5-form.
The leading $\a'$ corrections to the tree-level   IIB string
 effective action implied by the structure of the Green-Schwarz
 4-point
massless string scattering amplitude in the NS-NS sector
(i.e. depending on the metric,  dilaton and
the  2-form $B_2$) can be written
as \rf{\grw,\gs}
\eqn\acor{\eqalign{
 S_8 &= {1\over 2\kappa^2 }\int\ d^{10}x  \sqrt{-g}\  \L_8\ ,\cr
\L_8 &= c_1 \a'{}^3 \e^{-{3\ov 2}\phi}
\bigg(  t_8^{abcdefgh}t_8^{mnpqrstu} +
{1\ov 8} \ee_{10}^{abcdefghij}\ee_{10}^{mnpqrstu}{}_{ij}
\bigg) \overline{C}_{abmn}\overline{C}_{cdpq}
\overline{C}_{efrs}\overline{C}_{ghtu} \ ,}  }
where
$  c_1 = {\zeta (3)\ov 3\cdot 2^{11}} $ and
\eqn\Cbar{
\overline{C}_{ijkl} = C_{ijkl} +
{1\ov 2} \e^{-{1\ov 2}\phi}\left( \nabla H\right)_{ijkl} -
 {1\ov 4} \left( \nabla^2 \phi\right)_{ijkl}\ , }
\eqn\nabH{
\left( \nabla H\right)_{ij kl} =
\nabla_i H_{jkl}-\nabla_j H_{ikl}\ ,  }
\eqn\nabphi{
\left( \nabla^2 \phi\right)_{ijkl} =
g_{ik} \nabla_j \nabla_l \phi - g_{jk} \nabla_i \nabla_l \phi -
g_{il} \nabla_j \nabla_k \phi + g_{jl} \nabla_i \nabla_k\phi \ .}
Here $C^{hmnk}$ is the Weyl tensor,
$\ee_{10}$ is the totally antisymmetric
symbol (we use Minkowski notation for the metric, so that
$\ee_{10}\ee_{10}=-10!$), and the tensor $t_8$
is defined in \schwarz\
(it involves only $\delta$-symbols but not
$\epsilon_8$).

A few important clarifications are in order.
The 4-point on-shell string scattering   amplitude determines
only terms of 4-th order in expansion of \acor\ near flat space
(and  modulo equation of motion
terms not visible in on-shell amplitude).
In particular, the $\ep_{10} \ep_{10}$ structure  whose expansion
starts with terms of 5-th order in the fields is not fixed  by it.
Fortunately, complementary  information is provided \gz\  by  the
known
4-loop sigma model beta function \gwz\ that allows us to restore
the  covariant non-linear form of the $R^4$ correction to the action
(see also \mye).
 However, the sigma model calculation \gwz\
was  carried out only for $B_2=0$ \foot{The generalization
of the 4-loop beta function computation of \gwz\
to the $B_2 \not=0$ case  is very complicated
and was not done so far in full (see \refs{\dk,\jjj}
for some partial results).}
and shed no light on the
$H$-dependent terms. In Appendix A we extract new information about
such terms from the 5-point  amplitude
for antisymmetric tensors and gravitons.

As follows from \gz,
there exists a  scheme
in which the  metric and dilaton
dependent terms in the  $O(\a'{}^3)$
action  written in the string frame
are given by\foot{For some useful relations  between $R^4$ invariants
see  \rf{\roo,\pvw}.}
\eqn\sfr{
 S=- { 1 \ov 2 \k^2}
\int d^{10} x \sqrt {-G} \ e^{-2\p}
\ \big[ R  + 4(\del \phi)^2  +  \a'{}^3 c_1 J_0 \big] \ , }
\eqn\jnu{ J_0 
= 3 \cdot 2^8 (R^{hmnk} R_{pmnq} R_{h}^{\ rsp} R^{q}_{\ rsk}
\ + \ {1\over 2}  R^{hkmn} R_{pqmn} R_h^{\ rsp} R^{q}_{\ rsk} )
}
$$ 
=(t_8 t_8 + { 1 \ov 8} \ep_{10} \ep_{10}) RRRR  
+ O(R_{mn}) \ .
$$
It is the action that 
follows from \sfr\ upon transformation to the Einstein frame
that is the correct  metric-dilaton action 
beyond  the 4-point order.
The terms linear in the second derivative
of the  dilaton in \Cbar\ may then be understood
as originating  from conformal transformation
to the Einstein frame $G_{mn} = e^{\p/2} g_{mn}$.

Due to the field redefinition ambiguity
\rf{\ts,\grw},     we can assume that all the terms in
\acor\ depend only on
the Weyl tensor, and there is no explicit dependence on the Ricci
tensor. Only such a choice of the $\a'{}^3$ corrections  is
directly compatible with the AdS/CFT correspondence (see also
\rf{\bg,\gkt}).\foot{Introducing
Ricci-tensor dependent terms would imply
that one would need a compensating field redefinition,
i.e. a change of a scheme.}

Ignoring  the derivatives of the
dilaton and the  3-form, eq. \acor\ may be
written as
\eqn\W{
\L_8 ={1\ov 8}\a'{}^3\zeta (3)\e^{-{3\ov 2}\phi} W  \ ,
\ \ \ \ \ \
W=  C^{hmnk}C_{pmnq}C_h{}^{rsp}C^q{}_{rsk}+
{1\ov 2}C^{hkmn}C_{pqmn}C_h{}^{rsp}C^q{}_{rsk}  \ .
}
To obtain \W\ one should use the well-known
symmetries of the Weyl tensor.
The tensor
$\nabla H$ in \nabH\
does not possess all of  these symmetries, in particular, it is
antisymmetric under the interchange of the first and second
pairs of indices:
$\left( \nabla H\right)_{ijkl}=-\left( \nabla H\right)_{klij}$.
For this reason, a priori
we are not allowed to use \W\  with $C \to \bC$
if the
NS-NS 2-form does not vanish.

As follows from world-sheet parity considerations,
 the NS-NS part of tree-level type II string effective action
must be even in $B_2$;   in particular,  it  should  not
contain terms linear in  $H_3=dB_2$.
\foot{The  $hhhB_2$ 4-point amplitude corresponding
to \acor\ indeed vanishes.}
 Thus,
to find leading corrections to  the
backgrounds with  vanishing $H_3$ we  may ignore
 all $H_3$-dependent terms in the effective action.

The tensor
$\nabla^2\phi$ in \nabphi\
possesses all the symmetries of  the Weyl tensor but
it is not traceless. Thus, if the dilaton
does not vanish but the NS-NS 2-form does, the
expressions \acor\ and \W\ differ by the trace terms originating from
the term $\ee_{10}\cdot \ee_{10} \bC^4$. These trace terms
can be readily taken into account in all the cases we shall  consider.

The R-R scalar $C_0$ and the R-R 3-form $F_3$ can be easily included
into the action \acor\ by using the $SL(2,{\bf Z})$-invariance of
type IIB string theory.
The  $SL(2,{\bf Z})$-invariant form of
 the complete   4-point  ($\a'^3 R^4 +...$) effective action
(including non-perturbative corrections)
which depends on
all of the type IIB  massless bosonic
fields except  the 5-form was proposed in \kp.
 In particular,
to account for the contribution
of $F_3$ one can just add to the  $\bC$ in \Cbar\  the term
\eqn\cad{
{1\ov 2}\e^{{1\ov 2}\phi} g_s \left( \nabla F\right)_{ijkl}\ ,  }
obtained from the $\nabla H$ term in \Cbar\
 by the change $\phi\to -\phi,\ \ H\to g_s F$.
We  shall  not explicitly
 include the terms depending on  $C_0$ in the effective
action: the action \twob\ is  quadratic in $C_0$,\foot{
The leading-order (supergravity)  $C_0 H_3\cdot  F_3$ term vanishes on
the backgrounds of \rf{ \KW , \KT}.}
and $C_0$  is trivial
for  the backgrounds \rf{ \KW , \KT}
we are studying; therefore, $C_0$-dependent terms
cannot affect the equations of motion for other fields.

The quartic effective action  may also
contain terms dependent on  $F_5$,
e.g., $C^2 (\nabla F_5)^2, $  $  (\nabla H_3)^2(\nabla F_5)^2,$
$(\nabla F_3)^2(\nabla F)^2$, $(\nabla F_5)^4$, etc., which
 are not known at present.
In what follows
we will assume that  such
 terms do not change the form of the leading
corrections to the backgrounds we are studying.

The 8-derivative term in the
effective action may contain also other non-linear
structures  that contribute
to the S-matrix only starting at the 5-point or higher level.
In particular, it was conjectured in \kp\ that
there are the  5-point $R^3 H^2$ terms of the form
\eqn\rrrhh{
\left( t_8t_8+{1\ov 8}\ee_{10}\ee_{10}\right) RRR\
\overline{H}_3^{2}\ ,
}
where
\eqn\baH{
\left(\overline{H}_3^2\right)_{ijkl}=
H_{ikm}H_{jlm}- H_{jkm}H_{ilm}\ }
effectively replaces one of the four factors of $R_{ijkl}$
in the $R^4$ term.
As  we will show
in Appendix A,
 these particular
 terms are actually
absent  from  the effective action.
This is an important consequence of the supersymmetry of the theory.
There is, nevertheless, the  term of the form \rrrhh\ with
$\overline{H}_3^{2}$ \baH\  replaced by a
different contraction of the two $H$-tensors
\eqn\Hi{
\left(H_3^2\right)_{ijkl}= H_{ijm}H_{klm}\ .}
This term was found as a contribution to the 1-loop
string  effective action
using  the covariant  NSR
formalism in  \pwv, and its presence is
demonstrated through a light-cone gauge
calculation in Appendix A.

We shall analyze  the background values of
 the  $\a'^3$ correction
terms in the NS-NS sector in Appendix B, and
show that they  behave in accord with the
conjectured
duality between the supergravity  background of \rf{ \KT, \KS}
and the
$SU(N+M)\times SU(N)$ gauge theory.

\newsec{Absence of  corrections to $AdS_5\times T^{1,1}$ background }

In this section we study the $AdS_5\times T^{1,1}$ background
of type IIB supergravity \refs{\rom,\KW}, which has constant dilaton and the metric
given by
\eqn\adst{
ds_{10}^2 = {L^2\over z^2} (dz^2 + dx_n dx_n) + L^2 ds^2_{T^{1,1}}
\ .
}
With proper normalization included
($L^4= { 27 \ov 4}  \pi g_s N \a'^2$)
the 5-form field strength is given by   \HKO\
$$ F_5 = \F_5 + *\F_5\ ,\ \ \ \ \ \ \ \ \ \ \
\F_5 = 27\pi\a'^2  N \ {\rm vol} (T^{1,1}) \ , $$
while the 3-forms are not turned on. To investigate possible leading
higher derivative corrections to this background it is sufficient to
analyze the terms of the form
$C^4, C^3\nabla^2\phi$ in the effective action \acor.
As was explained in the previous section, all other
terms,   containing, e.g.,  the 3-form field strengths
$H_3$ and $F_3$, are at least quadratic in the
fields, and they cannot affect the equations
of motion in leading-order
perturbation theory.

By substituting the Weyl tensor for the metric \adst\
into the  $C^4$  invariant $W$ in \W\ we have checked
by a computer calculation that it vanishes. This means that the
term  $e^{-3\phi/2} W$
does not induce a dilaton tadpole.
 We further need to check that
there is no dilaton tadpole induced by the terms of the form
$e^{-3\phi/2} C^3 \nabla^2 \phi$. Luckily,
such sources for the dilaton vanish
as well. We will perform their calculation for the KT solution in
Appendix B;
the $AdS_5\times T^{1,1}$ result may then be extracted
in the  limit  $M\rightarrow 0$.

Let us note that the metric \adst\ can
be brought by a Weyl transformation
to the form
\eqn\dirr{
 (ds^2_{10})'= dx_n dx_n + dz^2 + z^2 ds^2_{T^{1,1}}\ ,
}
which describes the direct product
of the flat space $R^{3,1}$   and  the conifold.
The latter is a  Ricci flat
space
of $SU(3)$ holonomy, i.e. its Weyl tensor has a vanishing eigenvalue.
This fact is related to the vanishing of the invariant $W$ \W.\foot{This
is  related to the fact that the
 on-shell   $\N=1,$ $ D=10$   superinvariant  \rf{\NT,\bg}
$\int  d^{10} x d^{16} \theta \  \Phi^4  \to
\int d^{10} x d^{16} \theta \ (\bar \theta  \gamma^{mnk}
\theta\bar \theta  \gamma^{pq}_{\ \ \  k}
\theta  R_{mnpq})^4$
depends  only on the Weyl tensor   (because of the identity  $
 \gamma^{mnk} \theta\bar  \theta \gamma_{mnl}\theta
\equiv 0$)  and is proportional to $W$.}
Since $W$ transforms under a Weyl
transformation by an overall rescaling,
its vanishing for the conifold   implies its vanishing for
$AdS_5\times T^{1,1}$.

There are good reasons to believe that the conifold corresponds to
an exact 2-d  CFT, i.e. in contrast to generic CY spaces \NS\
it survives all $\alpha'$ corrections without deformation  of
 its metric.
Evidence for
this is provided by the linear sigma model formulation which defines
the conifold as an exact tree level string background \Witten.
Equipped with the explicit form of
the leading  higher-derivative correction
to the effective action
we can check a weaker statement: that the conifold
survives a leading order perturbation in $\alpha'$.
We have already mentioned  that $W$
vanishes when evaluated on this background.
It only remains to check that the same is true for
$\delta W/\delta g_{mn}$, so that
the Einstein equation continues to be
satisfied in leading order perturbation theory. We have checked
by a computer calculation that this is indeed  true.
A related statement is that the $E_6=\ep_6 \ep_6 RRR$
(6-d Euler density) correction
 \rf{\gwz,\pof}
to the Kahler potential (and also to the dilaton)
 of the ${\cal N}=2$ sigma model
for the conifold
vanishes  as well.
Interestingly, this is no longer the case for the resolved and
deformed conifolds.\foot{While $W$ still
vanishes  for the resolved and deformed
conifolds (as, in fact, for any similar 
space of special holonomy), 
its variation and the cubic  $E_6$ invariant
 do not, implying that the metric and dilaton
receive  $\a'^3$-corrections (note that while 
the $R^4$ term in the dilaton equation vanishes, 
there is an extra  correction term $D^2 R^3$ that does not). 
One heuristic argument
indicating why the singular
conifold
is not deformed by $\a'$-corrections  is based on the fact  that
the corresponding constrained  sigma model
$L = { 1 \ov 4 \pi \a'}  [ \del^a z_i \del_a z^*_i
+ (\Lambda z^2_i + c.c )] + $fermions
(here $i=1,..., 4$ and $\Lambda$ is Lagrange multiplier field)
has no intrinsic scale, i.e. is homogeneous
and quadratic in $z_i$.
Thus the dependence on $\a'$
can be absorbed into  $z_i \to \sqrt{\a'} z_i$.}

Similarly, we have checked that
$\delta W/\delta g_{mn}$ vanishes for the $AdS_5\times T^{1,1}$
space.
This establishes that the leading
perturbation in $\alpha'$ does not
change the form of this background. A similar check for the
$AdS_5\times S^5$ background is trivial as there each factor
 of the Weyl tensor  vanishes separately \bg.\foot{
A different  argument for stability of
$AdS_5\times S^5$  based on its maximal supersymmetry
was suggested in \KR.}
In the present  $AdS_5\times T^{1,1}$   case,
the Weyl tensor is non-vanishing so that the
direct confirmation of $\delta W/\delta g_{mn}=0$  was  necessary.

 This conclusion is quite important
 from the point of view of the
AdS/CFT correspondence. The $SU(N)\times SU(N)$ gauge theory
dual to the type IIB $AdS_5\times T^{1,1}$
background is conformal for
all $N$, and for all values of the gauge couplings.
The relation between the gauge couplings and
the moduli of string theory
is \refs{\KW,\MP,\KN,\HKO}
\eqn\edii{
{4\pi^2\ov g_1^2} + {4\pi^2\ov g_2^2} = {\pi\ov g_s}\e^{-\phi}\ ,}
\eqn\ediii{
\left({4\pi^2\ov g_1^2} - {4\pi^2\ov g_2^2}\right)
g_s\e^{\phi}= {1\ov 2\pi\a'} \int_{S^2}\ B_2\
-\pi \ \ \ ( {\rm mod}\  2\pi)\ .}
If $W$ were non-vanishing, then it would induce
a radial variation of the
dilaton which would have to be interpreted
as being due to
 a non-vanishing  beta function
in the field theory. This would be in
conflict with the vanishing of NSVZ
beta functions for both gauge couplings.
Thus, the fact that the $AdS_5\times T^{1,1}$ background survives
the leading perturbation in
$ \a'{}^3/L^6 \sim (g_s N)^{-3/2}$ is a new check of the
correspondence with the field theory
where
$g_s N$ is a modulus of the CFT.
We see that the AdS/CFT duality requires
that all $\alpha'$ corrections
and all string loop ($1/N$) corrections vanish for this background.
Proving this is quite a challenge. Perhaps
the vanishing of all $\alpha'$
corrections for this metric is related to the ability to
Weyl rescale it to the direct product \dirr\ of
 $R^{3,1}$ and  the conifold  and use the
fact that the conifold is an exact solution of string theory,
as well as to the
supersymmetry (with eight supercharges) of this background \rom.

\newsec{Fractional 3-branes on the conifold
  and RG flow of couplings}

In this section we proceed to the more complicated
case of the cascading
theory \refs{\KT,\KS}
which,  at the {\it bottom} of the cascade,  has gauge group
$SU(N+M)\times SU(N)$, $0\leq N < M$.
For $N=0$ the gravity dual of this theory
is given by the solution
of \rf{\KS}, while for $N>0$ it is  represented
 by an  appropriate
generalization which includes $N$ additional D3-branes on
the deformed conifold.
In practice, we will only consider the asymptotic UV (large radius)
form of these backgrounds derived in \KT; this
asymptotic form encodes the logarithmic RG flow that may be compared
with the NSVZ beta functions of the gauge theory.

First we review the solution of \KT,  complete with the normalization
factors supplied in \HKO.
The 10-d metric is
\eqn\metr{
ds_{10}^2=h^{-1/2}(r) dx_ndx_n+ h^{1/2}(r)(dr^2+r^2ds^2_{T^{1,1}})\ ,
}
where
\eqn\hhh{
h(r)= {27\pi\a'^2  \ov 4r^4} \
\big[ g_s N +
{ 3 \ov 2 \pi} (g_s M)^2 (  \lg (r/\tilde r)
+ {1 \ov 4}  ) \big] \ .
}
The 5-form is given by
\eqn\wedd{ F_5 =dC_4 + B_2 \wedge F_3=  \F_5 + *\F_5\ ,\ }
\eqn\ruu{
\F_5 = 27\pi\a'^2 \bN_{\rm eff}(r){\rm vol}(T^{1,1})\ , }
where (cf. \hhh)
\eqn\nenn{\bN_{\rm eff}(r) \equiv  N + {3\ov 2\pi }g_s M^2
\lg (r/\tilde r)\  . }
The 3-form field strengths are determined by
\eqn\gf{F_3 = {M\alpha'\over 2}\omega_3\ ,  \qquad\qquad
B_2 = {3 g_s M \alpha'\over 2}\omega_2 \log (r/\tilde r)
\ ,}
$$
H_3 = dB_2 = {3 g_s M \alpha' \over 2r} dr\wedge \omega_2\ ,
$$
where $\omega_2$ and $\omega_3$ are the harmonic forms on $T^{1,1}$
given in \HKO.
This background \KT\
provides a nice illustration of the distinction between  the
gauge-invariant but not localized and not quantized  charge  defined by
the generalized field strength,
$${1\over (2\pi)^4 \a'^2}
\int_{T^{1,1}} F_5 = \bN_{\rm eff}(r)\ ,  $$
and the quantized charge
\eqn\chau{
{1\over (2\pi)^4 \a'^2}
\int_{T^{1,1}} dC_4=
{1\over (2\pi)^4 \a'^2}\int_{T^{1,1}} (F_5 - B_2 \wedge F_3) =
N \ .  }
The latter is the analog of the Page charge (see, e.g.,  \mar).
It is  quantized because a probe D3-brane couples directly to $C_4$;
however, it is not invariant under the global gauge
transformations of $B_2$ --
 it is defined  modulo  integer shifts,
$ N \to N + k M$. This  is a reflection of the
duality cascade jumps \KS\
and, in fact, is a general phenomenon in a  system
with  different types of  fluxes.

One can rewrite $h(r)$ in the form
\eqn\hehh{
h(r)={\rL^4\ov r^4}\lg (r/r_s),\ \ \ \ \ \ \
\ \rL^2 \equiv {9g_s M\a'\ov 2\sqrt{2}}  \ . }
Performing the following change of coordinates
 \eqn\change{
t = \lg (r/r_s)\ ,\ \  \ \ \ \ \ \   \tx_n = {r_s\ov \rL^2}x_n\  ,
}
we can  put the  metric and $B_2$ into the form
\eqn\metr{
ds_{10}^2=\rL^2 \big[ {\e^{2t}\ov \sqrt{t}} d\tx_nd\tx_n+
\sqrt{t}(dt^2+ds^2_{T^{1,1}})\big]\ , }
\eqn\Bf{
B_2 ={3\ov 2} g_s M \a' \left(t - \tilde t \right)\omega_2\ ,}
where
\eqn\too{
\tilde t =\log (\tilde r/r_s)={2\pi N\ov 3g_s M^2}+ {1\ov 4}\ .}
Note that the scalar curvature is
$$ R = {2\ov \rL^2 t^{3/2}}\ ,$$
and no matter how small $\rL^2/\a'\sim g_s M$ is,
$\a' R$ can be made very small at large $t$.
Thus, corrections to supergravity can be organized in inverse
powers of $g_s M$ and $t$.

For $N=0$ we recover the asymptotic large distance ($\tau$) form of the
KS solution \KS; in this limit $\tau \rightarrow 3 t +{1\over 4}$.
Note that $\bN_{\rm eff} (t)$ can be written
in terms of $\rL$ and $g_s$ as
\eqn\nen{
\bN_{\rm eff}(t) = {4 \rL^4\ov 27\pi g_s\a'^2}\left(
t-{1\ov 4}\right)={3 g_s M^2\ov 2\pi}\left(
t-{1\ov 4}\right)\ .
}
Since $F_3$ and $H_3$ can be  also  expressed in terms
of $\rL$ (or $M$) and $g_s$,
and ${1\ov 2\pi\a'}\int_{S^2} B_2$ is an angular
variable  that  takes values in the interval $[0,2\pi],$
the solution has no explicit dependence on $N$.

The gravitational background describes the RG cascade of the
$SU(N+(k+1)M)\times SU(N+kM)\equiv
SU(N_{\rm eff}+M)\times SU(N_{\rm eff})$
gauge theories.
The $SU(N_{\rm eff}+M)\times SU(N_{\rm eff})$ theory
has the following symmetry
\eqn\symm{
M \to -M\ ,\ \ \ N_{\rm eff} \to N_{\rm eff} + M\ ,\ \ \ \ \ \
g_1\to g_2\ ,\ \ g_2\to g_1\ .}
The combination
 \eqn\brr{\bN = N_{\rm eff} + {M\ov 2}   }
is invariant under this
transformation, and, therefore, it is natural to
organize the expansion in powers of $1/(g_s \bN)$ and $M/\bN$.
Let us  define the point $\bt$ by the equation
\eqn\defbt{
\bN_{\rm eff}(\bt)=N + {M\ov 2}\ ,\ \ \ \ \ \ \ \
\bt = \tilde t +{\pi\ov 3g_s M} =
{2\pi\ov 3g_s M^2}\left( N +{M\ov 2}\right) +{1\ov 4}\ .}
Then in the vicinity of the point
$t_{\rm eff} = \bt + k {2\pi\ov 3 g_s M}$ such that
\eqn\defteff{
\bN_{\rm eff}(t_{\rm eff})=N + k M + {M\ov 2}
=N_{\rm eff} + {M\ov 2}=\bN\  ,}
the gravity background describes
the $SU(N_{\rm eff}+M)\times SU(N_{\rm eff})$
gauge theory.\foot{With our definitions, $N_{\rm eff}$
is automatically an integer. On the other hand,
$\bN_{\rm eff} (r)$ in \nenn\ continuously varies with $r$.
One may wonder how this continuous variation is consistent
with the statement that the number of colors makes discrete
jumps only at certain radii. We believe that
$\bN_{\rm eff} (r)$ can actually be interpreted as a measure
of the number of degrees of freedom for all $r$. This
is supported,  for example, by the smooth temperature dependence
of the Bekenstein-Hawking entropy for black holes embedded
in the asymptotically KT geometry \bhkzt.
The logarithmic scale dependence of the effective number of colors in
 between the
cascade jumps is presumably due to the interaction effects in
the $SU(N_{\rm eff}+M)\times SU(N_{\rm eff})$
gauge theory. It  would be interesting to study it directly in the
gauge theory.
}
The
background field strengths
(curvatures)
 are small provided that
$g_s \bN \gg 1$.

Following \refs{\KN,\KS,\HKO} we will
now use the relations \edii,\ediii\
to extract the supergravity prediction for the scale dependence of
the gauge couplings. Thus, we assume
that \edii,\ediii\ are valid not only
for constant dilaton and $B_2$, but also
when they are radially varying.
Substituting the $B_2$ from \gf\ into \ediii\ we find that \HKO
\eqn\flow{
{4\pi^2\ov g_1^2} - {4\pi^2\ov g_2^2} = 3 M (t-\bt )\
  \ \ ({\rm mod}\  {2\pi\ov g_s})\ \ ,  \ \ \
\qquad
{4\pi^2\ov g_1^2} + {4\pi^2\ov g_2^2} = {\pi\over g_s}\ .}
Thus, any point $t_{\rm eff}$, in
particular, $\bt$, is also
characterized by the requirement that the running gauge couplings
$g_1(t)$ and $g_2(t)$  obey  $g_1(t_{\rm eff})=g_2(t_{\rm eff})$.
In fact, due to the symmetry
\symm\ even if we include all possible corrections
to the dilaton and $B_2$,
there always exists such a point $t_{\rm eff}$ that
$g_1(t_{\rm eff} )=g_2( t_{\rm eff})$. It is not difficult to check
by using \flow\ that
at the point $t_1=t_{\rm eff} - {\pi\ov 3 g_s M}$
the gauge coupling $g_1$
of the bigger gauge group diverges, and that
at the point $t_2=t_{\rm eff} + {\pi\ov 3 g_s M}$
the gauge coupling $g_2$
of the smaller gauge group does.
These are the points where the
RG cascade jumps occur.

Let us emphasize that the supergravity description
is valid for large
$g_s N_{\rm eff}$ even if $g_s M$ is very small.
The separation between the cascade steps is $\Delta t = {2\pi\over
3 g_s M}$. Thus, there is a range of parameters where the cascade
jumps are far from each other, and
the supergravity  calculation of the
$\beta$-function
\flow\ may be compared with the NSVZ
$\beta$-functions for the $SU(N_{\rm eff}+M)\times SU(N_{\rm eff})$
theory:\foot{Here by the  NSVZ $\beta$-functions
we actually mean the SV  \shiv\ $\beta$-functions
for the holomorphic gauge couplings  which differ from the NSVZ ones
by  the 
absense of the denominator factor (this point was 
implicit in \KN\ and was  already 
 mentioned  \HKO).  
Indeed, in comparing to the dual string/supergravity
description one  should be using Wilsonian  effective 
action on the gauge  theory side. With the 
``holomorphic'' definition of 
gauge coupling as a factor in front of the 
$\N=1$ supersymmetric kinetic term 
 the corresponding $\beta$-function  is effectively ``one-loop''
 \shiv, i.e. does not have the denominator factor.
 This  
 definition of the coupling is indeed 
 consistent with the one based on 
 D-brane probe action and interpreted 
 as a quantum
gauge theory effective  
action defined at some fixed scale.
Let us mention  also that the  original 
expression \shif,\shiv\ 
for the NSVZ or SV    $\beta$-functions  does apply 
to the case of the non-semisimple  gauge group 
like the one  in the present example.
We are grateful to M. Shifman for a clarifying 
discussion of these  points. 
}
\eqn\nsvzi{
{d\ov d\log( \La / \mu )}\left( {4\pi^2\ov g_1^2} +
{4\pi^2\ov g_2^2}\right) = 2 \bN \D\ ,}
\eqn\nsvzii{
{d\ov d\log( \La /\mu )}\left( {4\pi^2\ov g_1^2} -
{4\pi^2\ov g_2^2}\right)= 3M - M\D \ ,}
where $\Delta$ is the correction to
the anomalous dimension $\g$
of the operators $\Tr A_iB_j$:
$$\g = -{1\ov 2} + \D \ .  $$
One expects $\D$ to scale as some even positive power of $M/\bN$ \KS.
Therefore, far in the UV where $\bN$ becomes
large due to the cascading phenomenon,
$\Delta$ naturally approaches zero.
Hence, the dominant running of the couplings in the UV is given by
\eqn\leadinglog{
{4\pi^2\ov g_1^2} - {4\pi^2\ov g_2^2}
\rightarrow 3M \log( \La / \mu )\ ,
\qquad
{4\pi^2\ov g_1^2} + {4\pi^2\ov g_2^2} \rightarrow {\rm const}
\ .
}
If we identify $t-t_{\rm eff}$ with $ \log( \La / \mu )$ then
the supergravity result \flow\ is in perfect agreement
with the field theory expectations.

The main purpose of this paper is to find out
where the effects of $\Delta$ are encoded on the string theory side.
We claim that these effects are provided precisely by the string
higher derivative corrections to the supergravity action.
In what follows we show that already the leading such correction
makes the dilaton radius-dependent, in agreement with \nsvzi\ which
implies that
$ {4\pi^2\ov g_1^2} + {4\pi^2\ov g_2^2}$ runs
due to the presence of $\Delta$.

\newsec{String correction to the KT solution and RG Flow}

To determine the leading $\a'{}^3$
correction to the dilaton
we have to take into account possible  mixing
between linearized fluctuations
of all of the relevant  supergravity modes.
In particular,  we should expect that not only the dilaton but
also the {\it 2-form $B_2$ } gets correction
at this order.

To find the mixing
we use the following ansatz for the deformed metric,
$B_2$  and dilaton:
\eqn\metran{
ds_{10}^2 = \rL^2 \bigg( {\e^{2t}\ov \sqrt{t}}\e^{2z} d\tx_n^2+
\sqrt{t}\e^{-2z}\left[\e^{10y} dt^2+ \e^{2y-8w}e_\psi^2+
\e^{2y+2w}\left(e_{\theta_1}^2+ e_{\theta_2}^2+
 e_{\phi_1}^2+e_{\phi_2}^2 \right)
\right]\bigg) \ , }
\eqn\Bfan{
B_2 = {3\ov 2} g_s M \a'[t+ b(t) ] \omega_2\ , \ \ \ \ \ \ \ \
\phi =\phi (t) \ .
}
Here the basis  $e_i$ is  the same as in \bhkzt.
The functions
$z,y,w,b,\phi$  depending only on the radial coordinate $t$
represent  the relevant fluctuations
around the KT solution. This ansatz preserves the symmetry
between the two spheres.

A few explanatory comments are in order.
It may seem  that the choice of the metric in \metran\ is too special.
In general, one may always  make a choice of radial coordinate $u$
so that a metric  with required symmetries  will be
\eqn\ran{
ds_{10}^2=\e^{2\hat z} d\tx_n^2+
\e^{-2\hat z}[\e^{10\hat y} du^2+ \e^{2\hat y-8\hat w}e_\psi^2+
\e^{2\hat y+2\hat w}\left(e_{\theta_1}^2+ e_{\theta_2}^2+
 e_{\phi_1}^2+e_{\phi_2}^2 \right)]\ . }
Here the three functions will then have expansion in powers of
$\a'$.  The metric \metran\ is  obtained from \ran\
by extracting the leading-order solution for these functions
and redefining $u\to t$. This implies  that $z,y,w$ in
\metran\  should start with $\a'{}^3$ correction terms.

We are assuming that $F_3$ is not modified and $F_5$ is
expressed through $B_2$ and $F_3$ by the standard  formula \wedd.
Indeed, it is easy to see that since  $F_3$ has purely magnetic
form,  the mixed ``Weyl tensor -- R-R 3-form''
term $C^2 (\nabla F_3)^2$ in the action \acor,\Cbar,\cad\
does not modify the equation for $C_2$ (the metric is diagonal
and has non-trivial dependence on $t$ only).
The same should be true
for the terms depending on $F_5$.

To find the  linearized equations of motion for the fluctuations
we need to find the quadratic action for the fluctuations
which follows from the type IIB supergravity  action
plus  terms linear in fluctuations
which follow from the
 $\a'{}^3$ string correction, i.e. from $ S_8$ \acor.
The computation of the leading  quadratic term in the  supergravity
Lagrangian \twob\  is straightforward (prime denotes
derivative  with respect to $t$)
\eqn\LL{\eqalign{
\L_2 = -{\rL^8 {\rm e}^{4 t}\over 2}
&\big[    \phi '{}^2  +  {4\ov t} {b}'{}^2 +
 40\ w '{}^2\ + 16\ {z}'{}^2\  - 40\ {y}'{}^2
-{8\ov t} {b}'\ \left(\phi +4w+4y-4z \right)\cr
&+{16\ov t^2} {b}^2
+ {4\ov t} \phi^2 +
  {32\ov t} \phi\  {w }
+ {64\ov t} w^2
+ 480\ w^2 +
  {32\ov t} {\phi}\ y +
  {128\ov t} {w }\ y  \cr
&+ {64\ov t} y^2 -
  1280\ y^2
- {64\ov t^2} b\ z +
  {256\ov t} b\ {{z}}
+ {32\ov t^2} z^2 -
  {192\ov t} z^2
+ 512\ z^2    \big]\ .  }}
Adding the $\a'{}^3$ correction term
$S_8\equiv {1\ov 2\kappa^{2}} \bar S_8 $ in \acor\
and taking the variational derivative
of  the action with respect to the fields, we derive
the following equations of motion for small perturbations:
 \eqn\eqb{\eqalign{
&{8 \rL^8 {\rm e}^{4 t}\over t}\Big( {1\ov 2} b'' + 2 b' -
{2\ov t} b' -{2\ov t} b - {1\ov 2}\phi'  + {1\ov 2t}\phi -
2\phi -2w'\cr
& -8w +  {2\ov t}w -2y' -8y +  {2\ov t}y +2z'-8z+
 {2\ov t}z \Big) + {\d \bar  S_8\ov \d b} = 0 \ ,}}
 \eqn\eqphi{\eqalign{
 \rL^8 {\rm e}^{4 t}\Big( \phi''+ 4\phi' - {4\ov t}\phi +
 {4\ov t} b' -{16\ov t}y -{16\ov t}w \Big)
+ {\d \bar  S_8\ov \d \phi} = 0 \ ,}}
 \eqn\eqy{\eqalign{
8 \rL^8 {\rm e}^{4 t}\Big( -5 y'' -20y'+160y-{8\ov t}y
 + {2\ov t} b' - {2\ov t}\phi - {8\ov t}w  \Big)
 + {\d \bar  S_8\ov \d y} = 0 \ ,}}
\eqn\eqz{\eqalign{
16 \rL^8 {\rm e}^{4 t}\Big(  z''+4z'-32z+  {12\ov t}z
-  {2\ov t^2}z  -  {1\ov t} b' -{8\ov t} b  +  {2\ov t^2} b
\Big) + {\d  \bar  S_8\ov \d z} = 0 \ ,}}
 \eqn\eqw{\eqalign{
8 \rL^8 {\rm e}^{4 t}\Big( 5 w'' +20w'-60w-{8\ov t}w
 + {2\ov t} b' - {2\ov t}\phi - {8\ov t}y  \Big)
 + {\d  \bar  S_8\ov \d w} = 0 \ .}}
Computing the $\a'{}^3$ terms to  linear order
in the fluctuations, we find
that all of  the variational derivatives have
the following leading behavior at
large $t$:
\eqn\varder{\eqalign{
{\d  \bar  S_8\ov \d \phi}  = 8\rL^8\e^{4t}{D_{\phi}\ov
t^{{7\ov 2}}}\ ,\ \  \ \ \ &
{\d  \bar  S_8\ov \d b}  = 8\rL^8\e^{4t}{D_{b}\ov
t^{{7\ov 2}}}\ ,\ \  \cr
{\d  \bar  S_8\ov \d y}  = 8\rL^8\e^{4t}{D_{y}\ov
t^{{5\ov 2}}}\ ,\ \ \ \
{\d  \bar  S_8\ov \d z}  =& 8\rL^8\e^{4t}{D_{z}\ov
t^{{5\ov 2}}}\ ,\ \ \ \
{\d  \bar  S_8\ov \d w}  = 8\rL^8\e^{4t}{D_{w}\ov
t^{{5\ov 2}}}\ ,\ \
}}
where $D_{\varphi_i}$ are some coefficients proportional
to $\a'{}^3/\rL^6$ whose exact values
are unknown due to the lack of
information about the terms involving
the 5-form $F_5$.
We present a detailed analysis of
these $\a'{}^3$ corrections
in Appendix B.

Comparing this
behavior with the equations of motion \eqb--\eqw,
we find that the leading
asymptotics  are
\eqn\asym{
b \sim {b_0\ov t^{{3\ov 2}}},\ \ \ \
\phi \sim {\phi_0\ov t^{{5\ov 2}}},\ \ \ \
z \sim {z_0\ov t^{{5\ov 2}}},\ \ \ \
y \sim {y_0\ov t^{{5\ov 2}}},\ \ \ \
w \sim {w_0\ov t^{{5\ov 2}}}\ .\ }
Let us pause and note that these
asymptotics are the minimal ones
compatible with the AdS/CFT correspondence between
string theory on the $AdS_5\times T^{1,1}$ background and the
$SU(N)\times SU(N)$ gauge theory.
Indeed,
the $AdS_5\times T^{1,1}$ background  \adst\
can be obtained from the KT one
\metr--\gf\
in the limit $M\to 0$ or, equivalently,
$\rL\to 0$.  More precisely,  one
should first redefine the variable $t$ as
\eqn\limi{
t = {2\pi N\ov 3g_s M^2} + \hat t  =
{\a'{}^2\ov \rL^4}{27\pi g_s N\ov 4} +\hat t \ ,}
and keep $\hat t $ fixed in the limit. Taking into account that
 the fluctuations are proportional to $\a'{}^3/\rL^6$, we find
that all of them except the function  $b$ vanish in the limit $\rL\to 0$.
While
$b$ itself goes to a constant, the  2-form $B_2$
vanishes in this  limit because of
the extra factor of  $M$ in \Bfan.
Thus  the $AdS_5\times T^{1,1}$ background is not modified,
in agreement with the discussion in Section 3. Note that if
some of the fluctuations (other than $b$)
would scale as $t^{-3/2}$,  the
$AdS_5\times T^{1,1}$  background  would acquire $\a'{}^3$
corrections.

Substituting the asymptotics \asym\ into
the equations of motion \eqb--\eqw,
we obtain
\eqn\eqbi{
-5 b_0 - 2 \phi_0
-8(z_0+y_0+w_0)  = - D_b \ ,}
 \eqn\eqphii{
-{3\ov 4}b_0 - {7\ov 4} \phi_0  - 2w_0-2y_0
 = - D_\phi  \ ,}
\eqn\eqzi{
-16 b_0 - 64 z_0 = - D_z \ ,}
\eqn\eqyi{
y_0  = - {D_y\ov 160}  \ ,\  \  \ \ \  \ \ \
 w_0  =  {D_w\ov 60} \ .}
Solving eqs. \eqbi--\eqzi, we find
\eqn\solphi{
\phi_0 = {4D_\phi\ov 5} -{D_b\ov 5}+{D_z\ov 40} \ , }
\eqn\solb{
b_0 = {7D_b\ov 15}-{8D_\phi\ov 15} -{2D_w\ov 45}-{7D_z\ov 120}
+{D_y\ov 60}\ ,
} \eqn\solz{
z_0 = {29D_z\ov 960}
-{7D_b\ov 60}+{2D_\phi\ov 15} +{D_w\ov 90}-{D_y\ov 240} \ . }
Let us now  use the results for
$\phi$ and $b$ to compute the correction $\D$ to
the anomalous dimension $\g = -{1\ov 2} + \D$
which enters the beta functions \nsvzi,\nsvzii.
Substituting \Bfan\ into \ediii, we  get
\eqn\idii{
{4\pi^2\ov g_1^2} + {4\pi^2\ov g_2^2} = {\pi\ov g_s}\e^{-\phi(t)}\ ,}
\eqn\idiii{
{4\pi^2\ov g_1^2} - {4\pi^2\ov g_2^2} =
\e^{-\phi(t)}\left( 3 M\left[ t-\bt + b(t)\right] \ \ \ \
({\rm mod} \ {2\pi\ov g_s})\ \right)
 .}
To compute the beta functions we differentiate \idii\
and \idiii\
over $t$.  Comparing to  \nsvzi\  we get
\eqn\reli{
\D = -{\pi\phi'\ov 2g_s \bN }\e^{-\phi}|_{t=t_{\rm eff}}
=-{2\pi^2\phi'\ov \l}\e^{-\phi}|_{t=t_{\rm eff}}\ ,       \ \ \
\ \ \ \l = 4\pi g_s \bN  \ .       }
This formula may also be written as
\eqn\relinew{
\D = -{2\pi^2\phi'\ov \l_{\rm eff} }\ , \qquad  \l_{\rm eff} = \l
\e^{\phi}|_{t=t_{\rm eff}}\ .}
Another expression for $\D$ is found from \nsvzii:
\eqn\relii{
\D = 3\e^{-\phi}\left(\e^{\phi}-1  -b'
+ b \phi'\right)|_{t=t_{\rm eff}} \ .}
To analyze these expressions
note that
\eqn\alphi{
{\a'\ov \rL^2}  = {2\sqrt{2}\ov 9 g_s M} =
{8\pi \sqrt{2}\ov 9 \l}\left(
{\bN\ov M}\right)\ ,\ \ \ \ \ \
{1\ov t_{\rm eff}}  \sim {3 g_s M^2\ov 2\pi \bN} =
{3\l \ov 8\pi^2}\left(
{M\ov \bN}\right)^2 .}
{}Since according to \asym, $\phi\sim \phi_0 t^{-5/2},\
b\sim b_0 t^{-3/2}$, where $\phi_0$ and $b_0$ are proportional to
$\a'{}^3/\rL^6$, we obtain
\eqn\phias{
\phi\sim b'\sim {\a'{}^3\ov \rL^6  t_{\rm eff}^{{5\ov 2}}} \sim
{1\ov \l^{{1\ov 2}}}\left({M\ov \bN}\right)^2 \ ,\ \ \ \ \ \ \
\phi'\sim {\a'{}^3\ov \rL^6 t_{\rm eff}^{{7\ov 2}}} \sim
 \l^{{1\ov 2}}\left({M\ov \bN}\right)^4 \ .}
Thus, by using \reli, we conclude that  at large $t_{\rm eff}$ and
at order $\a'{}^3$
\eqn\di{
\D \sim {1\ov \l^{{1\ov 2}}}\left({M\ov \bN}\right)^4\ .}

On the other hand, naively,  eq.\relii\ seems to give
a different expression
$
\D \sim {1\ov \l^{{1\ov 2}}}\left({M\ov \bN}\right)^2 .
$
A way to  avoid contradiction is to assume that
 at order $\a'{}^3$ the
fluctuations $b$ and $\phi$ satisfy the constraint
\eqn\conb{
b'(t) = \phi(t)\ .}
Then  eq.  \relii\  would imply the
absence of corrections to this order.
However, there are also higher order
 $\a'{}^5$ corrections to the dilaton and
$B_2$, and it is easy to see that assuming that
the  $\a'{}^5$ corrections scale at large $t$  in such a way that
\eqn\ses{
b'-\phi  \sim {\a'{}^5\ov t^{{9\ov 2}}}\ ,}
one obtains from \relii\ a correction $\D$ with the same
dependence on   $\l$ and $M/\bN$  as in \di.

Working to leading order in $M/\bN$ we may replace $\l$ in \di\
by $\l_{\rm eff}$. This latter expression is convenient
because \idii\ directly relates $\l_{\rm eff}$ to the gauge couplings.
Thus, we arrive at
\eqn\dcorrnew{\Delta
= a_1\left ({M\over \bN}\right )^4
\left ( {4\pi^2 \over \bN g_1^2} + {4\pi^2 \over \bN g_2^2}\right )^{1/2}
\ ,
}
where $a_1$ is proportional to
$c_1$ in \acor\ or $\zeta(3)$.
With this expression for $\Delta$, the $\beta$-function equation
\nsvzi\ assumes the form
\eqn\nsvznew{
{d\ov d\log( \La / \mu )}\left( {4\pi^2\ov \bN  g_1^2} +
{4\pi^2\ov \bN g_2^2}\right)^{1/2} =
a_1\left ({M\over \bN}\right )^4\ .}
Its solution  is
\eqn\nsvzsol{
\left( {4\pi^2\ov \bN g_1^2} +
{4\pi^2\ov \bN g_2^2}\right)^{1/2} = 2 \pi
\lambda_{\rm eff}^{-1/2} + a_1\left ({M\over \bN}\right )^4
\log( \La / \mu ) \ .}
Substituting this back into \dcorrnew\ we find
\eqn\dcorrscale{\Delta
= 2 \pi a_1\left ({M\over \bN}\right )^4
\lambda_{\rm eff}^{-1/2}+
O[(M/\bN)^8 \log( \La / \mu )]
\ .}
Finally, substituting this into \nsvzii\ and integrating we get
\eqn\nsvzsoli{
{4\pi^2\ov g_1^2} - {4\pi^2\ov g_2^2}
= 3M \log( \La / \mu ) \left ( 1 - {2\over 3}
 \pi a_1\left ({M\over \bN}\right )^4 \lambda_{\rm eff}^{-1/2}
+ O[({M/ \bN})^8 \log( \La / \mu ) ]\right )\ .
}
It is remarkable that the solutions of the NSVZ equations including
the effects of non-zero $\Delta$ have  an
expansion in powers of $\log( \La / \mu )$.
{}From the point of view of dual string theory this property of
RG flow is guaranteed by the relations \idii,\idiii,
and by the fact that
$t-t_{\rm eff}$  has to be identified with $\log( \La / \mu )$.
Moreover, the choice of the expansion point
  does not change the  logarithmic character of the expansion.
A closely related observation
is that
changing the variable according to \limi,
and expanding in  small $M$, or, equivalently, in small
$\hat t$
 (i.e. expanding near  $AdS_5 \times T^{1,1}$ background
dual to the  conformal fixed point of \KW)
we find that the $\a'$ corrections and the solutions
of the resulting  effective equations of motion  can be represented
in terms of power series in the logarithmic scaling variable
$\hat t$.
It would be very interesting to demonstrate the existence of
higher powers of $\log( \La / \mu )$ without invoking the
gauge field/string duality.

\newsec{Concluding Remarks}

In this paper we studied the leading stringy
corrections to two different
solutions of type IIB supergravity,
interpreting our results in terms of
dual gauge theories.
We were able to use the gauge/gravity duality in both directions,
sometimes using constraints from field theory as a way of predicting
properties of the string effective action.

For the $AdS_5 \times T^{1,1}$ background we demonstrated
that the $O(\alpha'^3)$ correction does not modify
the form of the solution,
in line with expectations from AdS/CFT duality.
In fact, turning the duality
around, we predicted a much stronger result: that the
$AdS_5 \times T^{1,1}$ background is exact to
all orders in $\alpha'$ and
$g_s$. We suspect that this property is
tied to the exactness of the
conifold background.

Our paper also contains a much more ambitious calculation of
leading string-theoretic corrections to the solution of \KT,
which describes
logarithmic RG flow and duality cascade in ${\cal N}=1$
supersymmetric $SU(N+M)\times SU(N)$ gauge theory.
This calculation is made especially difficult by the fact that the
complete structure of the $O(\alpha'^3)$ correction in the
(tree-level)
type IIB superstring effective action
is not yet known.
Nevertheless,  we are able to demonstrate an interesting interplay
between the NSVZ $\beta$-functions in field theory and the
structure of stringy corrections. In fact, we may again
turn the gauge/gravity duality around and use field theory
to predict certain facts about the string theory effective action.
The most basic fact is the absence from the effective
action of terms which scale as $1/t^3$ at large $t$.
If such a term were present,  then
agreement with the
gauge/gravity duality
would fail for this theory.
We were able to check this prediction
of gauge theory for string theory for all terms
in the effective action from the NS-NS sector.

There were also more subtle predictions that we
were unable to check completely, such as the relation
\conb\ at order $\a'{}^3$. The latter  implies
a relation between the coefficients
$D_{\varphi_i}$ in \varder:
$
D_b = {2\ov 15} D_w -{1\ov 20} D_y +{1\ov 8} D_z  .
$
We concluded also
that corrections to $\phi - b'$ at the order
$\alpha '{}^{5}$  should be  completely determined by corrections
to $\phi$ at order $\alpha '{}^{3}$.
We expect these relations
to be  consequences of the special
structure of the string  $\a'{}^3$ correction to the effective
action dictated by supersymmetry.
\foot{The fact
that there is only one anomalous dimension
that enters the two NSVZ beta functions
(r.h.s. of \nsvzi\ and \nsvzii)  implies
that there is a specific combination of the two
gauge couplings (or of the dilaton and 2-form $B_2$)
whose dependence on the scale (or $r$) is known exactly
to all orders in the $M/N$ expansion.
}

Assuming that these properties hold,
we are able to make a prediction
for the correction
$\Delta$ to the anomalous dimension of $\Tr (A_i B_j)$
in the gauge theory.
The result is consistent with the expectation \KS\
that in field theory $\Delta$ must
have an expansion in powers of $(M/\bN)^2$:
our  supergravity analysis
suggests that the leading term has the form \dcorrnew.
The  normalization factor $a_1$ in this formula
is proportional to $\zeta(3)$.
It would be very interesting, but probably  hard,
 to understand  this string-theory prediction
directly on the gauge theory side.

Some  methods developed in  this paper may have applications
in other related contexts.
One may perform a similar study of leading $\a'$ corrections
to the solutions considered in \gip.
Our result (in Appendix A) about the absence of
certain
$R^3 (H_3)^2$ terms in  the
$(R^4+...)$ superinvariant  in type IIB
  10-dimensional  effective action
may imply certain  constraints
on  possible  $R^3 (F_4)^2$ terms in the 11-dimensional
effective action,  and this may be important
in the context of  the discussions
in \rf{\TTT,\BB}.
Thus, a complete determination of the 8-derivative corrections
 to the
type II
supergravity action should have many useful applications.

\bigskip
\bigskip

\noindent
{\bf Acknowledgements}

Some of the calculations with the $AdS_5\times T^{1,1}$
background presented here
have been independently carried out by C.~P. Herzog.
We are grateful to him for useful discussions.
We are also grateful to Yu.N. Obukhov for help
with GRG computer algebra program used in our calculations.
We acknowledge  M. Green, I. Jack, K. Skenderis
and  especially
K. Peeters, P. Vanhove and A. Westerberg
for very  useful comments on the original version of the paper
that helped us to clarify and correct  the presentation.
The work of S.~F. and A.~A.~T.  was supported by
the DOE grant
DE-FG02-91ER40690.
The work of I.~R.~K. was supported in part by the NSF
grant PHY-9802484.
The work of A.~A.~T.  was also supported in part by
the  PPARC SPG 00613,
 INTAS  99-1590 and  CRDF RPI-2108 grants.

\appendix{A}{ $R^3 H^2$ terms in type II
tree level effective action}
The tree-level effective action of type IIB and type IIA
theories is the same in the NS-NS (metric, $B_2$ and dilaton)
sector.  This action  should be parity-even,
and, apart from the central charge
term, should have the  universal dimension-independent
structure.\foot{The corresponding NSR sigma model
can be formulated in any dimension $D$ and thus
its beta-functions must  have dimension-independent universal
coefficients (in a standard dimensional regularization with
minimal subtraction scheme).
This is no longer  so once
we include the dependence on the R-R fields which are sensitive
to $D=10$.}
  Using sigma-model considerations, it is easy to see
that the effective action should be even in $H_{mnk}$, i.e.
it may contain terms of the form
$R^3 H^2, \ R^2 H^4$, etc.
Our aim is  to study the possible presence of the  $R^3 H^2$ terms
in the effective action.
They may a priori   accompany  the  $R^4$ terms on dimensional
and  (non-linear) $D=10$ supersymmetry grounds.

In this Appendix we show that certain irreducible
5-point terms of the form
$R^n H^{5-n}$  are absent
from the type II superstring effective action.\foot{
Here we  call a term reducible if it can be represented as  a sum
of terms of the form
$R^{m}(\nabla H)^{n}$ by using $[\nabla ,\nabla ]H\sim R H$.
We shall consider $R^n H^{5-n}$  for general $n=1, ..., 5$
with understanding that, in view of the above remarks,
the only potentially non-trivial cases are $n=1,3$
(the $R^5$-terms, i.e. $n=0$,   are absent  as suggested  by
the result  of \kaz\ and also by the   $\N=2, D=10$
supersymmetry).
}
We shall directly  compute the corresponding
5-point terms in the effective action
using  the Green-Schwarz
light-cone formulation of the   superstring  S-matrix \grsch.

Since we shall use the light-cone gauge approach in its
standard most straightforward  form, i.e.
assuming that the components of
the polarization tensors vanish in the
two light-cone directions,
we shall not be able to  determine   a  special class of terms
that involve antisymmetrization of  {\it  nine} (or more)
 space-time indices, e.g.,
$\epsilon_{k m_1 ... m_{9}} \epsilon^{k n_1 ... n_{9}}
R^{m_1 m_2}_{n_1 n_2} R^{m_3 m_4}_{n_3 n_4}
 R^{m_5 m_6}_{n_5 n_6}
 H^{m_7}_{\ n_7 n_8}  H^{\ m_8 m_9}_{n_9}
$.\foot{We are grateful to
K. Peeters, P. Vanhove and A. Westerberg
for pointing out  to us the possible presence of such term
in the
(one-loop) type II effective action,  as
well as of the term
$t_8t_8 R^3 H^2$ which we missed in the first version of
this paper \pwv.}
However, it is possible to check that precisely because
such terms involve antisymmetrization of nine indices,
they do not change our conclusion about the $1/t^4$ scaling of the
leading $\a'^3$ corrections  to  our background.


In the notation of \gsw\ the light-cone gauge
vertex operators for the graviton $h_{ij}$
and the NS-NS 2-form $b_{ij}$ can
be written as ($i,j,...=1,...,8$)
\eqn\vgr{
V(h) = \left( h_{ik}\dXL^i \dXR^k
- {i\ov 4} \G_{ik,l} \dXL^i \tS\g^{kl}\tS
- {i\ov 4} \G_{ki,j} \dXR^k S\g^{ij}S
- {1\ov 32} R_{ijkl} S\g^{ij}S \tS\g^{kl}\tS\right)
\e^{i p X}
,}
\eqn\vb{
V(b) = \left( b_{ij}\dXL^i \dXR^j
+ {i\ov 8} H_{ikl} \dXL^i \tS\g^{kl}\tS
+ {i\ov 8} H_{kij} \dXR^k S\g^{ij}S
+ {i\ov 32} p_l H_{ijk} S\g^{ij}S \tS\g^{kl}\tS\right)
\e^{i p X}
.}
Here $ \G_{ij,k}(p) =
{i\ov 2}( p_i h_{jk}(p)+ p_j h_{ik}(p) - p_k h_{ij}(p))$
is a linearized Christoffel symbol,
$ R_{ijkl}(p) = -{1\ov 2}
( p_i p_l h_{jk}(p)-p_j p_l h_{ik}(p)
+ p_j p_k h_{il}(p) -p_i p_k h_{jl}(p))$ is a
linearized Riemann tensor, and
$H_{ijk}(p) = i( p_i b_{jk}(p)- p_j b_{ik}(p) + p_k b_{ij}(p))$ is
the NS-NS 3-form. We also denote $ \dXL^i \equiv \pa_- X_L^i,\
\dXR^k \equiv \pa_+ X_R^k$, and $\pa_\pm$ are derivatives
over the world-sheet  directions.

Our aim is to study the structure of some tree-level
 5-point massless scattering amplitudes involving
the graviton and 2-form field.
Following the logic similar to the one
in  \grw, we may use a  short-cut argument.
Due to the supersymmetry and the expected  $SL(2,{\bf Z})$ invariance
 of the ``massless''  type IIB superstring effective action,
for  $\a'{}^3 (R^4 + ...)$ terms in the tree-level
effective action there should
exist  terms of the same structure in the one-loop effective action
 \mgr.
To  get a nonzero one-loop amplitude
one should saturate the integral
over the fermionic zero modes -- 8 modes $S_0$, and 8 modes $\tS_0$.
Thus, the 5-point amplitude has to be given by a sum of
the following terms
\eqn\ampli{ \eqalign{
&A_5 \sim  V_{i_1j_1k_1l_1}  V_{i_2j_2k_2l_2}  V_{i_3j_3k_3l_3}
V_{i_4j_4k_4l_4} V_{i_5j_5k_5l_5}
(t)_{10}^{i_1j_1...i_5j_5}
(t)_{10}^{k_1l_1...k_5l_5}
\int dp
\Tr\left( \e^{i p_a X_a}w^{L_0}\bar{w}^{\tilde{L}_0} \right)
\cr
&+V_{i_1j_1k_1l_1}  V_{i_2j_2k_2l_2}  V_{i_3j_3k_3l_3}
V_{i_4j_4k_4l_4}   V_{i_5k_5}
(t)_{8}^{i_1j_1...i_4j_4}
(t)_{8}^{k_1l_1...k_4l_4}
\int dp
\Tr\left(
 \dXL^{i_5}\dXR^{k_5}\e^{i p_a X_a}
w^{L_0}\bar{w}^{\tilde{L}_0} \right)
\cr
&+ V_{i_1j_1k_1l_1}  V_{i_2j_2k_2l_2}  V_{i_3j_3k_3l_3}
V_{i_5 k_4l_4} V_{k_5i_4j_4}
(t)_{8}^{i_1j_1...i_4j_4}
(t)_{8}^{k_1l_1...k_4l_4}
\int dp
\Tr\left(
\dXL^{i_5}\dXR^{k_5}\e^{i p_a X_a}
w^{L_0}\bar{w}^{\tilde{L}_0} \right)
}}
Here $V_{ijkl}$ is either $R_{ijkl}$ or $p_l H_{ijk}$,
$V_{ikl}$ is either $\G_{ik,l}$ or $H_{ikl}$, and
$V_{ik}$ is either $h_{ik}$ or $b_{ik}$.
 The tensor
$(t)_{2m}$
is defined as  \gsw
\eqn\tens{
(t)_{2m}^{i_1j_1...i_{m}j_{m}} = 4^{-m}\tr
\left(S_0\g^{i_1j_1}S_0\cdots S_0\g^{i_mj_m}S_0 \right)\ ,}
where the trace is over the spinor zero modes of $S$.
One can show \gsw\ that $(t)_8= t_8 - \ha \ep_8$,
where $t_8$ is given by
 the sum of products of the $\delta$-symbols.

Since the light-cone set-up  is essentially
8-dimensional,
it does not actually allow  one to determine
 the presence of,  e.g., the
$\ep_8 \ep_8 RRRR$ term in the action
directly, since this term is a
total derivative in 8 dimensions to all orders in
the  graviton expansion.
In general, fixing
 the  8-derivative 5-point
terms depending on $\ep_8$, i.e.
 $\ep_8 \ep_8 R^3H^2$ or
$\ep_8 t_8 R^3 H^2$, is subtle 
 in the light-cone
formulation (in particular,
because of possible contact terms needed for
space-time supersymmetry \grs).
%
For this reason, we will only
determine the terms with the $t_8 t_8$ tensor structure.

It is understood  that the
trace of the operators
$\e^{i p_a X_a}w^{L_0}\bar{w}^{\tilde{L}_0}$ and
$ \dXL^{i}\dXR^{k}\e^{i p_a X_a}w^{L_0}\bar{w}^{\tilde{L}_0}$
is taken  over the
nonzero modes of $X_L,X_R,S,\tS$. The integration
over the positions of the 5 vertex
operators and the modulus parameter $w$ is implied.
   The integral
over the zero mode momentum $p$ can be regarded as the trace
over the zero modes.
It is  because of this  integral
that the
closed string amplitude can not be
treated simply as  a product of
the two open string ones.

It is clear that the first term in \ampli\ leads to
terms of the form $R^n (\nabla H)^{5-n}$ in the effective action.

To analyze the contribution of the second and third terms to
the 5-point amplitude we note that due to the $SO(8)$ invariance
\eqn\trdxdx{
\int d^{8}p
\Tr\left(
\dXL^{i}\dXR^{k}\e^{i p_a X_a}
w^{L_0}\bar{w}^{\tilde{L}_0} \right)
= p_a^{i} p_b^{k} A^{ab} +\d^{ik} D\ ,
}
where $A^{ab}, D$ are some functions of momenta $p_a$, positions
of the vertex operators and the modulus parameter $w$.
Note that the $\d^{ik}$ term appears only because of the
integration over the zero mode momentum $p$, and because
both $\dXL^{i}$ and $\dXR^{k}$ depend on $p$.

Now one can easily see that
the second term in \ampli\ cannot lead to a
term $R^nH^{5-n}$ since the
indices $i_5,k_5$ are contracted with $V_{i_5k_5}$.
Moreover, if $V_{i_5k_5} = b_{i_5k_5}$ then the second term
must be absent since
it is not invariant under the gauge transformation
$B_2 \to B_2 + d\zeta$ of the NS-NS 2-form while all other terms in
\ampli\ are.  If $V_{i_5k_5} = h_{i_5k_5}$ then we get
vanishing result   since, as usual, one assumes that
$h_m^m=0$ in the  on-shell vertex operator.

To analyze the third term in \ampli\ we need to consider
three cases. In all the cases the three vertices
$V_{ijkl}$ can be either $R_{ijkl}$ or $p_l H_{ijk}$.

$(i)$ $V_{i_5k_4l_4}=\G_{i_5k_4,l_4},\
V_{k_5i_4j_4}=\G_{k_5i_4,j_4}$.
The tensor $p^{i_5}_a p^{k_5}_b$ in \trdxdx\ leads to
the terms of the form
$R^n(\nabla H)^{3-n}\del \G \del \G$,\foot{We use
a shorthand notation $\del \G$ to denote
$\pa^k \G_{ki,j}$, where the derivative $\pa_k$ may act
on any of the fields in $R^n(\nabla H)^{3-n}\del \G \del \G$.}
and
because of the reparametrization invariance they can contribute
either to the covariantization of the
term $R^{n+1}(\nabla H)^{3-n}$ coming from the
4-point amplitude or to a term of the form
$R^{n+2}(\nabla H)^{3-n}$. The $\d^ {i_5k_5}$ term in \trdxdx\
gives $R^n(\nabla H)^{3-n}\G \G$ and contributes to
the covariantization of the
term $R^{n+1}(\nabla H)^{3-n}$.

$(ii)$ $V_{i_5k_4l_4}=H_{i_5k_4l_4},\
V_{k_5i_4j_4}=\G_{k_5i_4,j_4}$.
This term is of the form $R^n(\nabla H)^{4-n}\del \G$
or $R^n(\nabla H)^{3-n} H \G$, and
the reparametrization invariance again requires that it contributes
either to the term $R^{n}(\nabla H)^{4-n}$
or to a term of the form $R^{n+1}(\nabla H)^{4-n}$.

$(iii)$ $V_{i_5k_4l_4}=H_{i_5k_4l_4},\  V_{k_5i_4j_4}=H_{k_5i_4j_4}$.
Such  term in the 5-point amplitude is gauge and
reparametrization invariant.
The tensor $p^{i_5}_a p^{k_5}_b$ in \trdxdx\ leads to
the terms of the form $R^n(\nabla H)^{5-n}$.
However, we also find the contribution from the
$\d^ {i_5k_5}$ term in \trdxdx,
which is of the form $R^n(\nabla H)^{3-n} H^2$, where
the contraction
$
\left({H}^2\right)_{ijkl}= H_{ijm}H_{klm}$ is the same
as in \Hi.

We conclude that,
 contrary to the earlier expectation \kp,
 there is  no term of the form \rrrhh,\baH\ in the effective action
but there is still a  $t_8 t_8 R^3 H^2$
term with $H^2$ given by \Hi.
The presence of such a term is demonstrated using the NSR
formalism in the forthcoming paper
\pwv, to which we refer for the  details
of the covariant form of the $R^3 H^2$ part of the 1-loop type
IIB(A)  string effective action.

Using  similar considerations
it may be possible to  rule out most of the
higher-dimensional terms of the
form $R^n H^k$  where all $H_3$
factors are not covered by derivatives.\foot{Naively, one could
expect  \kp\   the presence of
$HH$ term \baH\ supplementing the curvature in each factor
 in \Cbar. This is
motivated by  the condition that the
corrections to equations of motion should vanish
for a group space (WZW model). However,
it is easy  to see that
sigma model perturbation theory does not imply
 that all corrections to beta-functions and thus
 also the effective action should depend on $H_3$ only through
 the curvature of the generalized connection
$\hat \Gamma = \Gamma \pm {1\ov 2}
H_3$ (see e.g. \rf{\mtt,\dk}).}

\appendix{B}{Structure  of $\a'{}^3$ corrections to KT solution}
To analyze $\a'{}^3$ corrections it is convenient to
use orthonormal zehnbeins $E_m$ (or  corresponding 1-forms)
 in terms of which our perturbed background
\metran,\Bfan\ may be written as
($n=1,2,3,4$, $\alpha=6,7,8,9$; we reserve index 0
for the radial direction $t$)
\eqn\metrn{
ds_{10}^2 = \rL^2 \big[ \e^{2z} E_n^2+
\e^{-2z}\left(\e^{10y} E_0^2+ \e^{2y-8w}E_5^2 +
\e^{2y+2w}E_\a^2
\right)\big]\ ,}
\eqn\Hft{
H_3 = {9g_s M\a'\ov 2 \rL^3}(1+b')
t^{-{3\ov 4}}E_0\wedge\left( E_6\wedge E_8 -
E_7\wedge E_9\right)
\ ,} \eqn\Fft{
F_3 = {9 M\a'\ov 2 \rL^3} t^{-{3\ov 4}}E_5\wedge\left( E_6\wedge E_8 -
E_7\wedge E_9\right)
\ .}
Our aim is to compute the correction \acor\
in this case.
We  assume that all indices in \acor\ are tangent.
A direct computation shows that the tensors appearing in
\Cbar\ have the following $t$-dependence:
\eqn\weyl{
C_{ijkl}\sim  C_{ijkl}^{(1/2)}t^{-1/2} + C_{ijkl}^{(3/2)}t^{-3/2}
+ C_{ijkl}^{(5/2)}t^{-5/2}\ ,}
where the coefficients $C_{ijkl}^{(1/2)}$ do not vanish only
if $i,j,k,l = 6,7,8,9$;
\eqn\nablaH{
\left( \nabla H\right)_{ijkl}\sim  h_{ijkl}^{(1)} t^{-1} +
h_{ijkl}^{(2)} t^{-2} \ ,}
where the coefficients $h_{ijkl}^{(1)}$ do not vanish only
if two of the indices $i,j,k,l$ take values from 6 to 9,
and the other two
indices take values from 1 to 6;
\eqn\nablaph{
\left(\nabla^2 \phi\right)_{ijkl}\sim  p_{ijkl}^{(1/2)} t^{-1/2} +
p_{ijkl}^{(3/2)} t^{-3/2} \ ,}
where the coefficients $p_{ijkl}^{(1/2)}$ do not vanish only
if at least two of the indices $i,j,k,l$ take values from 1 to 6.

This $t$-dependence shows that if there were no cancellations,
the $C^4$ and $C^3\nabla^2\phi$ terms
would scale at large $t$ as $1/t^2$, and the terms
$C^2(\nabla H)^2$,  $C(\nabla H)^2\nabla^2\phi$ and $C^3 H^2$
would scale as
$1/t^3$.

We will see, however, that due to a special structure of
these  $\a'^3$  terms (related to the supersymmetry of
the underlying IIB theory)
when evaluated on the KT solution they scale as
$1/t^4$. The first variations of these   terms with respect to the
fluctuations $y,z,w$ scale as $1/t^3$.
Taking into account that
\eqn\sqrtg{
\sqrt{-g}= \rL^{10}\e^{4t} \sqrt{t}\ \e^{10y-2z} \ ,}
we  find that the variational derivatives of $S_8$ are indeed given
by \varder.

The computation of the  invariant $W$ \W\ is
straightforward and gives
\eqn\w{
W={1\ov \rL^8}\big(   {40\ov t^4} - {36\ov t^5} +
{452\ov 27\,t^6}- {35\ov 6\,t^7} +
{85\ov 16\,t^8} - {75\ov 32\,t^9}  +
{225\ov 512\,t^{10}}      \big)\ .
}
Thus  at large $t$  we get $W \sim  {1\ov t^4}$.
This implies that
in the limit $\rL \to 0$,\  $t \sim 1/\rL^4$ leading to
the $AdS_5\times
T^{1,1}$ case
 the invariant $W$  {\it vanishes}.

It is also not difficult to compute the terms in $W$ which are
linear in fluctuations $y,z,w$
\eqn\CCCdC{
\d W = {1\ov \rL^8}\left(
{640 y\ov t^3}+ {288 y'\ov t^3}+{32 y''\ov t^3}-
 {320 z'\ov t^3}- {64z''\ov t^3}-{960w\ov t^3} -{192w'\ov t^3}
 -{48 w''\ov t^3}\right) \ ,}
where we  kept only  terms that are leading at large $t$.
These terms give contributions to the equations of motion
for $y,z,w$  which are  of order $t^{-5/2}$. Again,
in the limit $\rL \to 0$, $t \sim 1/\rL^4$ these  terms vanish.

Computing $C^3\nabla^2\phi$ by using the simplified formula \W,
we find
that all the terms with the $1/t^2$ scaling cancel.
There are  still two
terms that scale  as $1/t^3$
\eqn\snphi{
C^3\nabla^2\phi \sim  {1\ov t^3}\left( \phi'' + 4\phi' \right)
\ .}
It is easy to see, however, that being integrated
over $t$ with the measure $\sqrt{-g} = \rL^{10} \e^{4t}\sqrt{t}$,
these two  terms cancel each other, and the leading
contribution to the dilaton equation of motion
is of order $t^{-7/2}$, as it was the case for the  $C^4$ term.
\foot{One can easily see that the trace terms originating from
$\ee_{10}\cdot \ee_{10}C^3\nabla^2\phi$ are either proportional
to the linearized dilaton equation of motion,
and have the same form as
\snphi, or are a
total derivative, and do not contribute to the equation.}
Thus the $C^3\nabla^2\phi$ term
only changes the coefficient in
the r.h.s. of the dilaton equation.

To show that the terms $C^2 (\nabla H)^2$,
$C (\nabla H)^2 \nabla^2\phi$ and
$C^3 H^2$ do not contain terms of order $1/t^3$
we need to recall that the invariants formed from the $t_8$
and $\ee_{10}$ tensors can be written as follows
\eqn\tete{
X\equiv t_8 t_8 \bC^4 = 192 I_{41} + 384 I_{42} + 24 I_{43} +
12I_{44}-96I_{45}-96\tilde{I}_{45}-48I_{46}-48\tilde{I}_{46}\ ,}
\eqn\eeee{
{1\ov 8}Z\equiv -{1\ov 8}\ee_{10}\ee_{10}\bC^4
= 192 I'_{41} + 384 I'_{42} +
24 I'_{43} +
12I'_{44}-192I'_{45}+96I'_{46}-768A_7 +{1\ov 8}Z'\ .}
Here $Z'$ are trace  terms, and
the fundamental
invariants are defined as \eqn\invrs{\eqalign{
&I_{41} = \tr \left(\bC_{mn}\bC_{nr}\bC_{rs}\bC_{sm}\right)\ , \cr
&I_{42} = \tr \left(\bC_{mn}\bC_{nr}\bC_{ms}\bC_{sr}\right)\ , \cr
&I_{43} = \tr \left(\bC_{mn}\bC_{rs}\right)
\tr \left(\bC_{mn}\bC_{rs}\right)\ , \cr
&I_{44} = \tr \left(\bC_{mn}\bC_{mn}\right)
\tr \left(\bC_{rs}\bC_{rs}\right)\ , \cr
&I_{45} = \tr \left(\bC_{mn}\bC_{nr}\right)
\tr \left(\bC_{rs}\bC_{sm}\right)\ , \cr
&I_{46} = \tr \left(\bC_{mn}\bC_{rs}\right)
\tr \left(\bC_{mr}\bC_{ns}\right)\ , \cr
&Z= \bC_{mn}{}^{[mn}\bC_{pq}{}^{pq}\bC_{rs}{}^{rs}\bC_{tu}{}^{tu]}\
,\cr
&A_7 =\bC_{pq}{}^{rs}\bC_{ru}{}^{pt}\bC_{tv}{}^{qw}\bC_{sw}{}^{uv}
\ .}}
The matrices $\bC_{mn}$ are naturally defined by
$(\bC_{mn})^a{}_b \equiv \bC_{mn}{}^{ab}$; the invariants
$\tilde{I}_{ab}$ are defined by the same
formulas with the replacement
$\bC_{mn}\to \bC_{mn}^T$, where
$(\bC_{mn}^T)^a{}_b = \bC_{ab}{}^{mn}$;
the invariants
$I'_{ab}$ are defined by replacing in
the formulas \invrs\ the
second and the fourth tensors $\bC$ with $\bC^T$.
When $\bC = \bC^T$ the above relations
reduce to the ones given in
\rf{\roo,\pvw}.

First,  we discuss the term  $C^2 (\nabla H)^2$.
Since we want to show
that it does not contribute to the order $1/t^3$ it is
sufficient to consider the terms in $C_{ijkl}$
that scale as $t^{-1/2}$,
and the terms in $\left(\nabla H\right)_{ijkl}$
that scale as $t^{-1}$. As was mentioned above,  in this case
$C_{ijkl}$ do not vanish only if $i,j,k,l = 6,7,8,9$,
and  $\left(\nabla H\right)_{ijkl}$ do not vanish only
if two of the indices $i,j,k,l$ take values from 6 to 9,
and the other two indices take values from 1 to 6. Moreover,
in any of the pairs $i,j$ or $k,l$ one of the indices
takes  values from 6 to 9,
and the other takes values from 1 to 6. For this reason,
we find that the $\ee_{10} \ee_{10}$ term
vanishes. The $t_8 t_8$ term has to be computed by using \tete,
and also gives zero. It is worth noting that in this consideration
we have not assumed that the fluctuation $b$ of $B_2$ vanishes.
Thus, this  term gives contributions of order $t^{-7/2}$ to both
the equations of motion for the $\phi$  and $b$.

To  show
that $C (\nabla H)^2 \nabla^2\phi$
does not contribute to the order $1/t^3$, we take $C_{ijkl}$ and
$\left(\nabla H\right)_{ijkl}$ with the same scaling as above, and
$\nabla^2\phi$ scaling as $t^{-1/2}$. Then a straightforward
computation shows that among the invariants
$I_{ab}$ only  $I_{41}=I'_{41}$ does not vanish.
This  invariant, however, does not appear in the difference
$X - {1\ov 8}Z$ that gives the $\a'{}^3$ correction.
Therefore, we need to show the
vanishing $-768 A_7\ +$ ``trace terms" in \eeee.
Computing $A_7$ and the ``trace terms", we get
\eqn\Asev{
A_7 = -{8\ov t^3} (4\phi' - \phi'')\ , \ \ \ \ \ \
{1\ov 8}Z' = -{8\cdot 768\ov t^3} (12\phi' + \phi'')\ . }
Thus, their total  contribution is
\eqn\chhp{
768 A_7 - {1\ov 8}Z' =  {16\cdot 768\ov t^3} (4\phi' + \phi'') \ .}
As was explained above,  it leads to
a contribution of order $t^{-7/2}$ to
the dilaton equation of motion. This is of the same
order as the contribution due to the $C^4$ term.

In \pwv\ and in Appendix A it was shown that
there are certain terms of
the form $(t_8t_8 + {1\ov 8}\ee_{10}\ee_{10})C^3H^2$ where
$H^2$ is given by \Hi. It is again clear that the
$\ee_{10} \ee_{10}$ term cannot contribute to the $1/t^3$ order.
Computing the $t_8t_8$ term by using \tete\ and \invrs,
we find that, although all the invariants $I_{4n}$ contribute
to the order $1/t^3$, the total contribution of
the $t_8t_8$ term starts with $1/t^4$. \foot{Though we have
proved in Appendix A that the term \rrrhh\ is not present in
the effective action, we have
checked that it also starts with $1/t^4$.}

Thus, we see that special properties of the 8-derivative corrections
to the type IIB effective action lead to a
number of cancellations when
evaluated on the KT background. These cancellations are needed for
consistency with the RG flow in the dual field theory.

\vfill\eject
\listrefs
\end

\\
Title: String Corrections to the Holographic RG Flow of
Supersymmetric SU(N) x SU(N+M) Gauge Theory
Authors:  Sergey Frolov, Igor R.~Klebanov and Arkady A.~Tseytlin
Comments: harvmac, 29 p.
Report-no: OHSTPY-HEP-T-01-024, PUPT-2004
\\
We study leading string corrections to the type IIB
supergravity solution dual to the
${\cal N}=1$ supersymmetric $SU(N+M)\times
SU(N)$ gauge theory coupled to bifundamental
chiral superfields $A_i, B_j$, $i,j=1,2$.
This solution was found in hep-th/0007191, and its asymptotic form
describing logarithmic RG flow was constructed in
hep-th/0002159. The leading  tree-level string correction
to the type IIB  string effective action is represented by
the invariant of the form $\a'{}^3 ( R^4 + ...)$.
Since the background contains 3-form
field strengths, we need to know  parts of this invariant
that depend on them.  By analyzing the 5-point superstring
scattering amplitudes
we show that only a few specific  $R^3 (H_3)^2$ and $R^3 (F_3)^2$
terms
are present in the effective action. Their contribution to
the holographic RG flow turns out to be of the same order as of
the $R^4$ terms.
This fact is crucial for finding
agreement between the $\alpha'$-corrected radial dependence of the
supergravity fields  and the RG flow dictated by the
NSVZ beta
functions in field theory.
The agreement with field theory
requires that the anomalous dimension
of the  operators $\Tr (A_i B_j)$ is corrected by a term of order
$(M/N)^4 \lambda^{-1/2}$ from its value $-{1 \over 2} $ found for $M=0$
($\lambda$ is the appropriate 't Hooft coupling).
\\